\documentclass[a4,11pt]{article}
\usepackage{amsmath,amsthm,amssymb,bm,mathrsfs}
\usepackage{comment}
\usepackage[dvipdfmx]{graphicx}
\usepackage{color}

\begin{document}

\renewcommand{\thefootnote}{\fnsymbol{footnote}}

\begin{center}
 {\Large \bf Boltzmann equation and its cosmological applications}

  \vspace{3ex}

  {\large Seishi Enomoto}\footnote{seishi@mail.sysu.edu.cn},
  {\large Yu-Hang Su},
  {\large Man-Zhu Zheng},
  {\large and Hong-Hao Zhang}\footnote{zhh98@mail.sysu.edu.cn}
  
  \vspace{2ex}
  
  {\it School of Physics, Sun Yat-sen University, Guangzhou 510275, China}
  
  \vspace{2ex}
  
  

\end{center}

\renewcommand{\thefootnote}{\arabic{footnote}}
\setcounter{footnote}{0}

\begin{abstract}


We review the derivation of the Boltzmann equation and its cosmological applications in this paper.  The derivation of the Boltzmann equation, especially the collision term, is discussed in detail in the language of the quantum field theory without any assumption of the finite temperature system.  We also discuss the integrated Boltzmann equation with the deal of the temperature parameter as an extension of the standard equation.  Among a number of its cosmological applications, we mainly target two familiar examples, the dynamics of the dark matter abundance through the freeze-out/in process and a baryogenesis scenario.  The formulations in those systems are briefly discussed with techniques in their calculations.
\end{abstract}

\section{Introduction}
The early Universe, composed of hot plasma, evolves in time while maintaining thermal equilibrium at most epochs, and thus the evolution is described simply by thermodynamics.  The cosmologically important events, therefore, are focused on the various turning epochs in which some particle species depart from the thermal bath as seen in, e.g., Big Bang nucleosynthesis (BBN) and recombination confirmed by both the theory and observations, and the dark matter (DM) relic abundance and the baryogenesis scenario derived from some hypotheses.  Qualitatively, the turning epochs can be estimated by comparing the reaction rate and the spatial expansion rate of the Universe, but more quantitative treatment to ensure accurate predictions is required by the present cosmological observations.
The Boltzmann equation is a powerful tool for following the out-of-equilibrium dynamics in detail, and can describe the evolution of the particle distribution due to the spatial expansion and the momentum exchanges through interactions.  Since the Boltzmann equation provides the abstract relation between the macroscopic and the microscopic evolution of the particle distributions, it is applicable to various situations, and hence, the solving equations differ for each system.

The most successful application of the Boltzmann equation is for the theory of BBN \cite{Alpher:1948ve,Hayashi:1950lqo,Alpher:1953zz,Burbidge:1958,Hoyle:1964zz,Zeldovich:1965gev,Peebles:1966rol} (see also, e.g., \cite{Kolb:1990vq,Weinberg:2008zzc,Cyburt:2015mya}), which describes how the initial protons and neutrons form the light elements at the final.  As the result, the abundances of the light elements are predicted, and they are well consistent with the present observations.  The other well-motivated and formulated example is the Kompaneets equation \cite{Kompaneets:1956,Weymann:1965}, which is derived from the Boltzmann equation in the photon-electron system.  The evolution equation is applied to the various analysis in cosmological or astronomical situations, e.g., the last scattering surface in the recombination epoch, the Sunyaev-Zel'dovich effect \cite{Sunyaev:1969,Zeldovich:1969ff,Sunyaev:1970,Sunyaev:1972} that is a distortion of the cosmic microwave background radiation by hot electrons in galaxies, etc.


On the other side, exploring Beyond the Standard Model (BSM), there are many studies on the DM candidate realizing the present density parameter through the relic abundance, which can be predicted using the Boltzmann equation.  The most famous and simple scenario for DM candidates is described by weakly interacting massive particles (WIMPs) \cite{Lee:1977ua,Scherrer:1985zt}, but the current observation requires a more extended scenario or the other mechanism, e.g., feebly interacting massive particles (FIMPs) \cite{Hall:2009bx}, strongly interacting massive particles (SIMPs) \cite{Hochberg:2014dra}, processing into forbidden channels \cite{Griest:1990kh,DAgnolo:2015ujb}, co-annihilations \cite{Griest:1990kh}, co-scattering \cite{DAgnolo:2017dbv,Kim:2019udq}, zombie \cite{Berlin:2017ife,Kramer:2020sbb}, and inverse decays \cite{Garny:2017rxs,Frumkin:2021zng}, etc.  (See also \cite{Frumkin:2022ror} for their summarized analysis.)  The more complicated models are considered, the more technical treatment in the Boltzmann equation is required.  As the other topic for exploring BSM with the Boltzmann equation, there are also many studies for baryogenesis scenarios explaining the origin of (baryonic) matter-antimatter asymmetry in our Universe, e.g., GUT baryogenesis \cite{Yoshimura:1978ex,Weinberg:1979bt,Dimopoulos:1978kv}, leptogenesis \cite{Fukugita:1986hr}, electroweak baryogenesis \cite{Cohen:1990py,Rubakov:1996vz}, and Affleck-Dine baryogenesis \cite{Affleck:1984fy,Dine:1995kz}, etc.  The analysis tends to be more complicated than the case of DM abundance and requires more technical treatment.

The studies using the Boltzmann equation have been more popular and important as some approaches to confirm the current physics in detail and explore BSM in the early Universe.  We aim to provide a clearer understanding of the Boltzmann equation and its techniques with some cosmological examples.  Although there is a more accurate and complicated formulation beyond the Boltzmann equation, the so-called Kadanoff-Baym eqaution \cite{Baym:1961zz,Baym:1962sx}, we do not deal with it in detail in this paper.  For readers who want to apply to the Kadanoff-Baym equation, see section \ref{sec:BE_limitation} and references there.

This paper is organized as follows.  First, we derive the Boltzmann equation by the language of the quantum field theory in section 2.  Next, we demonstrate application examples how the Boltzmann equation is applied for the DM abundance in section 3 and the baryogenesis scenario in section 4 with simple toy models.  Finally, we summarize our discussion in section 5.

\section{Boltzmann equation}
The Boltzmann equation is a quite powerful tool to describe the evolution of particles in Cosmology.  Although the equation is applicable to various situations, the used formulae are also dependent on the circumstance.  In this section, we derive the evolution formula of the distribution function $f=f(x^\mu,p^\mu)$ from the basic statement
\begin{equation}
 \mathbf{L}[f]=\mathbf{C}[f]
\end{equation}  
where $\mathbf{L}$ and $\mathbf{C}$ are called the Liouville operator and the collision operator, respectively.  The Liouville operator describes the variation of the distribution of a particle along a dynamical parameter, and the collision operator describes the source of the variation through the microscopic processes.

\subsection{Liouville operator}
We define the Liouville operator to describe the variation of the distribution along the geodesic line parametrized by the affine parameter $\lambda$.  Using the momentum relations
\begin{eqnarray}
 p^\mu=\frac{dx^\mu}{d\lambda}, \qquad p^\mu p_\mu=m^2
\end{eqnarray}
where $p^\mu$ is a four-momentum and $m$ is the mass, and the geodesic equation
\begin{eqnarray}
 0=\frac{dp^\mu}{d\lambda}+\Gamma^\mu_{\nu \rho}p^\nu p^\rho
\end{eqnarray}
where $\Gamma^\mu_{\nu\rho}$ is the affine connection, the Liouville operator can be written as
\begin{eqnarray}
 \mathbf{L}[f] \:\: \equiv \:\: \left.\frac{df}{d\lambda}\right|_{\rm geodesic \ line}
  &=& \frac{dx^\mu}{d\lambda}\partial_\mu f+\frac{dp^\mu}{d\lambda}\frac{\partial f}{\partial p^\mu} \\
  &=& p^\mu \partial_\mu f-\Gamma^\mu_{\nu \rho}p^\nu p^\rho \frac{\partial f}{\partial p^\mu}.
\end{eqnarray}

In the FLRW space-time, $g_{\mu\nu}={\rm diag}(1,-a^2,-a^2,-a^2)$ where $a=a(t)$ is the scale factor, the distribution function is spatially homogeneous and isotropic: $f=f(t,E)$.  Then the concrete representation of the Liouville operator term can be written as
\begin{eqnarray}
 \mathbf{L}[f]=E\dot{f}-H|\vec{p}|^2\frac{\partial f}{\partial E} \label{eq:Boltzmann_geodesic}
\end{eqnarray}
where $H=\dot{a}/a$ is the Hubble parameter and $(\vec{p})^i\equiv ap^i$ is the physical momentum.

\subsection{Collision operator}
We define the collision operator $\mathbf{C}[f]$ as the variation rate by the microscopic processes:
\begin{eqnarray}
 \mathbf{C}[f] \:\: \equiv \:\: \left.\frac{df}{d\lambda}\right|_{\rm microscopic \ process} \:\: = \:\: \left. E \: \frac{df}{dt}\right|_{\rm microscopic \ process}. \label{eq:Boltzmann_collision0}
\end{eqnarray}
In order to evaluate the variation of the distribution function, we assume the followings.  First, the process can be described by the quantum field theory.  Second, the microscopic process can be evaluated on the Minkowski space since the gravitational effect is already evaluated in the Liouville operator part.

With the above assumption, the distribution function $f$ can be regarded as the expectation value of all possible occupation numbers with their probabilities, as we will see later.  Also the probabilities can be evaluated by the quantum field theory on the flat space.  To derive the concrete representation of (\ref{eq:Boltzmann_collision0}), we need to construct the corresponding quantum state and then evaluate the transition probability.

\subsubsection{Eigenstate for occupation number}
At first, we construct a multi-particle state $|\{n\}\rangle$ in order to include all information about the particle occupations.  We impose $|\{n\}\rangle$ to be the eigenstate satisfying
\begin{eqnarray}
 \hat{n}_a(\vec{k})|\{n\}\rangle=n_a(\vec{k})|\{n\}\rangle, \qquad \langle \{n\}|\{n\}\rangle=1,
\end{eqnarray}
where $\hat{n}_a(\vec{k})$ and $n_a(\vec{k})$ are the occupation operator and its corresponding occupation number for species $a\in\{n\}$ in the unit phase space, respectively.  Here the occupation operator is defined by 
\begin{equation}
 \hat{n}_a(\vec{k})\equiv\frac{1}{V}a_{\vec{k}}^{(a)\dagger} a_{\vec{k}}^{(a)},
\end{equation}
where $V\equiv \int d^3x=(2\pi)^3\delta^3(\vec{k}=0)$ is a volume of the system, and $a_{\vec{k}}^{(a)}$ is an annihilation operator for species $a$ which satisfies
\begin{equation}
 [a_{\vec{k}}^{(a)},a_{\vec{p}}^{(b)\dagger}]=\delta^{ab}\cdot (2\pi)^3\delta^3(\vec{k}-\vec{p}),\qquad ({\rm others}) = 0.
\end{equation}
In the case of the fermionic species, the commutation relations are replaced with the anti-commutation relations.  Then one can obtain the representation of the eigenstate $|\{n\}\rangle$ by
\begin{eqnarray}
 |\{n\}\rangle \equiv \prod_{a\in \{n\}}\left(\prod_{\vec{p}}\frac{1}{\sqrt{n_a!}\cdot \sqrt{V^{n_a}}}\:(a_{\vec{p}}^{(a)\dagger})^{n_a}\right)|0\rangle. \label{eq:eigenstate_for_occupation}
\end{eqnarray}
Note that the occupation number $n_a(\vec{k})$ must be an integer.

Furthermore, it is convenient to define the increased/decreased state from $|\{n\}\rangle$ for later discussion.  We define them by\footnote{For the bosonic state, the $N$-increased/decreased state can be defined by
\begin{eqnarray}
 |\{n\};\vec{k}_a^{(+N)}\rangle
  &=& \sqrt{\frac{n_a!}{(n_a+N)!}\frac{1}{V^N}} \cdot (a_{\vec{k}}^{(a)\dagger})^N |\{n\}\rangle, \\
 |\{n\};\vec{k}_a^{(-N)}\rangle
  &=& \sqrt{\frac{(n_a-1)!}{(n_a+N-1)!}\frac{1}{V^N}} \cdot (a_{\vec{k}}^{(a)})^N |\{n\}\rangle.
\end{eqnarray}
}
\begin{equation}
 |\{n\};\vec{k}_a^{(+1)}\rangle=\frac{1}{\sqrt{1\pm n_a}\sqrt{V}} \: a_{\vec{k}}^{(a)\dagger} |\{n\}\rangle,\qquad \left( \ +: {\rm bosons}, \quad -: {\rm fermions} \ \right),
\end{equation}
\begin{equation}
 |\{n\};\vec{k}_a^{(-1)}\rangle=\frac{1}{\sqrt{n_a}\sqrt{V}} \: a_{\vec{k}}^{(a)} |\{n\}\rangle.
\end{equation}
The coefficients are chosen to be unit vectors
\begin{equation}
 \langle \{n\};\vec{k}_a^{(\pm 1)} |\{n\};\vec{k}_a^{(\pm 1)}\rangle=1.
\end{equation}
These increased/decreased states also become the eigenstate of the occupation operator:
\begin{equation}
 \hat{n}_a(\vec{k})|\{n\};\vec{k}_a^{(+1)}\rangle=\left(1\pm n_a(\vec{k})\right)|\{n\};\vec{k}_a^{(+1)}\rangle,\qquad \left( \ +: {\rm bosons}, \quad -: {\rm fermions} \ \right),
\end{equation}
\begin{equation}
 \hat{n}_a(\vec{k})|\{n\};\vec{k}_a^{(-1)}\rangle=\pm\left(n_a(\vec{k})-1\right)|\{n\};\vec{k}_a^{(-1)}\rangle,\qquad \left( \ +: {\rm bosons}, \quad -: {\rm fermions} \ \right).
\end{equation}

\subsubsection{Transition probability}
Using the eigenstates discussed in the previous section, let us consider the transition probability of the process
\begin{eqnarray}
 A,B, \cdots \rightarrow X,Y,\cdots \label{eq:target_process}
\end{eqnarray}
in the background in which other particles ($\{n\}$) exist. For simplicity, we consider a case that each species in the process are different.  Taking the initial state as $|\{n\}\rangle$ in order to begin the given occupation numbers, the final state through the process (\ref{eq:target_process}) can be represented as $|\{n\};\vec{k}_A^{(-1)},\vec{k}_B^{(-1)},\cdots, \vec{k}_X^{(+1)},\vec{k}_Y^{(+1)},\cdots\rangle$.  The probability from the infinite past (in-state) to the infinite future (out-state) on the background particles can be evaluated by\footnote{If the initial or the final state includes $N$ of the same species, the extra factor $\frac{1}{N!}$ for each duplicated species is needed.
}
\begin{eqnarray}
 \mathcal{P}(A,B,\cdots\rightarrow X,Y,\cdots)_{\{n\}}
  &=& \prod_{a=A,B,\cdots,X,Y,\cdots} \left(V\int \frac{d^3\vec{k}_a}{(2\pi)^3}\sum_{g_a}\right) \nonumber \\
  & & \times \left|\langle \{n\};\vec{k}_A^{(-1)},\vec{k}_B^{(-1)},\cdots, \vec{k}_X^{(+1)},\vec{k}_Y^{(+1)},\cdots|\hat{S}|\{n\}\rangle\right|^2 \label{eq:def_probability}
\end{eqnarray}
where $g_a$ denotes the internal degrees of freedom for species $a$, and $\hat{S}$ is the S-matrix operator.
The S-matrix element describing the process (\ref{eq:target_process}) without the background particles can be represented by the invariant scattering amplitude as
\begin{eqnarray}
 {}_{\rm inv}\langle k_X,k_Y,\cdots|\hat{S}|k_A,k_B,\cdots\rangle_{\rm inv}
  &=& i\mathcal{M}(k_A,k_B,\cdots\rightarrow k_X,k_Y,\cdots) \nonumber \\
  & & \quad \times (2\pi)^4\delta^4(k_A+k_B+\cdots-k_X-k_Y-\cdots) \label{eq:S-M-relation}
\end{eqnarray}
where
\begin{eqnarray}
 |k_a,k_b,\cdots\rangle_{\rm inv}
  \equiv \sqrt{2E_{k_a}2E_{k_b}\cdots}\:\:a_{\vec{k}_a}^\dagger a_{\vec{k}_b}^\dagger \cdots |0\rangle
\end{eqnarray}
is a Lorentz invariant particle state.  The representation (\ref{eq:S-M-relation}) indicates that the S-matrix operator includes
\begin{eqnarray}
 \hat{S} &\supset& \prod_{a=A,B,\cdots,X,Y,\cdots}\left(\int \frac{d^3\vec{k}_a'}{(2\pi)^3}\frac{1}{\sqrt{2E_a'}}\sum_{g_a}\right)\times a_{\vec{k}_X'}^\dagger a_{\vec{k}_Y'}^\dagger \cdots a_{\vec{k}_A'} a_{\vec{k}_B'}\cdots \nonumber \\
 & & \quad \times i\mathcal{M}(k_A',k_B',\cdots\rightarrow k_X',k_Y',\cdots)\cdot (2\pi)^4 \delta^4(k_A'+k_B'+\cdots-k_X'-k_Y'-\cdots).
\end{eqnarray}
Using the above expression, the S-matrix element on the particle background can be written as
\begin{eqnarray}
 & & \langle \{n\};\vec{k}_A^{(-1)},\vec{k}_B^{(-1)},\cdots \vec{k}_X^{(+1)},\vec{k}_Y^{(+1)},\cdots|\hat{S}|\{n\}\rangle \\
 & & = \:\: \prod_{a=A,B,\cdots,X,Y,\cdots}\left(\int \frac{d^3\vec{k}_a'}{(2\pi)^3}\frac{1}{\sqrt{2E_A'}}\sum_{g_a}\right) \nonumber \\
 & & \quad \times i\mathcal{M}(k_A',k_B',\cdots\rightarrow k_X',k_Y',\cdots)\cdot (2\pi)^4 \delta^4(k_A'+k_B'+\cdots-k_X'-k_Y'-\cdots) \nonumber \\
 & & \quad \times \langle \{n\};\vec{k}_A^{(-1)},\vec{k}_B^{(-1)},\cdots \vec{k}_X^{(+1)},\vec{k}_Y^{(+1)},\cdots|a_{\vec{k}_X'}^\dagger a_{\vec{k}_Y'}^\dagger \cdots a_{\vec{k}_A'} a_{\vec{k}_B'}\cdots|\{n\}\rangle \\
 & & = \:\: \frac{1}{\sqrt{2E_AV\cdot 2E_BV\cdot \cdots 2E_XV \cdot 2E_YV\cdot \cdots}} \nonumber \\
 & & \quad \times i\mathcal{M}(k_A,k_B,\cdots\rightarrow k_X,k_Y,\cdots)\cdot (2\pi)^4 \delta^4(k_A+k_B+\cdots-k_X-k_Y-\cdots) \nonumber \\
 & & \quad \times \sqrt{n_An_B\cdots (1\pm n_X)(1\pm n_Y)\cdots }
\end{eqnarray}
where $+/-$ is for bosonic/fermionic particles of the produced species $X, Y, \cdots$. 
As substituting the above form into (\ref{eq:def_probability}), one can obtain
\begin{eqnarray}
 \mathcal{P}(A,B,\cdots\rightarrow X,Y,\cdots)_{\{n\}}
  &=& \prod_{a=A,B\cdots,X,Y,\cdots} \left(\int \frac{d^3\vec{k}_a}{(2\pi)^3}\frac{1}{2E_{k_a}}\sum_{g_a}\right) \nonumber \\
  & & \quad \times \left|\mathcal{M}(k_A,k_B,\cdots\rightarrow k_X,k_Y,\cdots)\right|^2 \nonumber \\
  & & \quad \times (2\pi)^4\delta^4(k_A+k_B+\cdots - k_X -k_Y-\cdots)\cdot VT \nonumber \\
  & & \quad \times n_An_B\cdots (1\pm n_X)(1\pm n_Y)\cdots \label{eq:total_probability}
\end{eqnarray}
where $T=\int dt=2\pi\delta(E=0)$ is the transition time scale.

Although the expression of the probability is derived, the result (\ref{eq:total_probability}) is constructed by the exact information of quanta represented by the microscopic occupation numbers per a unit phase space $n_a$ that is an integer.  Since it is impossible to know the exact quantum state, the statistical average should be considered.   The probability to realize the state $|\{n\}\rangle$ can be represented by
\begin{equation}
 P_{\{n\}}\equiv \prod_{a\in\{n\}}\prod_{\vec{k}} p(n_a(\vec{k})), \qquad \sum_{n_a=0}^\infty p(n_a(\vec{k}))=1 \label{eq:probability_for_statics}
\end{equation}
where $p(n_a(\vec{k}))$ is a probability to be the occupation $n_a$ on the momentum $\vec{k}$.
Multiplying (\ref{eq:probability_for_statics}) into (\ref{eq:total_probability}) and summing over by each occupation numbers\footnote{This procedure is equivalent to
\begin{equation}
 |\{n\}\rangle\langle \{n\}| \rightarrow \sum_{\{n\}}P_{\{n\}}|\{n\}\rangle\langle \{n\}|
\end{equation}
in (\ref{eq:def_probability}), that is, the initial state is considered by the density operator.}, we can obtain the statistical probability as
\begin{eqnarray}
 \langle \mathcal{P}(A,B,\cdots\rightarrow X,Y,\cdots)_{\{n\}}\rangle
  &\equiv& \sum_{\{n\}} P_{\{n\}}\cdot \mathcal{P}(A,B,\cdots\rightarrow X,Y,\cdots)_{\{n\}} \nonumber \\
  &=& \prod_{a=A,B,\cdots,X,Y,\cdots} \left(\int \frac{d^3\vec{k}_a}{(2\pi)^3}\frac{1}{2E_a}\sum_{g_a}\right) \nonumber \\
  & & \quad \times \left|\mathcal{M}(k_A,k_B,\cdots\rightarrow k_X,k_Y,\cdots)\right|^2 \nonumber \\
  & & \quad \times (2\pi)^4\delta^4(k_A+k_B+\cdots - k_X -k_Y-\cdots)\cdot VT \nonumber \\
  & & \quad \times f_Af_B\cdots (1\pm f_X)(1\pm f_Y)\cdots \label{eq:total_probability2}
\end{eqnarray}
where we denoted
\begin{equation}
 f_a\equiv \sum_{n_a=0}^\infty p(n_a(\vec{k})) \: n_a(\vec{k}).
\end{equation}
The important thing is that the expectation value $f_a$ can be interpreted as the distribution function even though the exact forms of both the probability $p(n_a(\vec{k}))$ and the relating occupation $n_a(\vec{k})$ are unknown.  

Using the result of the total probability (\ref{eq:total_probability2}), one can also define the partial probability, as an example, for the species $A$ of the momentum $\vec{k}_A$ by
\begin{eqnarray}
 \mathit{p}_A(A,B,\cdots\rightarrow X,Y,\cdots)
  &\equiv& \frac{d\langle \mathcal{P}(A,B,\cdots\rightarrow X,Y,\cdots)_{\{n\}}\rangle}{\displaystyle V\frac{d^3\vec{k}_A}{(2\pi)^3}\sum_{g_A}} \\
  &=& \frac{T}{2E_A}\prod_{a\neq A} \left(\int \frac{d^3\vec{k}_a}{(2\pi)^3}\frac{1}{2E_a}\sum_{g_a}\right) \nonumber \\
  & & \quad \times \left|\mathcal{M}(k_A,k_B,\cdots\rightarrow k_X,k_Y,\cdots)\right|^2 \nonumber \\
  & & \quad \times (2\pi)^4\delta^4(k_A+k_B+\cdots - k_X -k_Y-\cdots) \nonumber \\
  & & \quad \times f_Af_B\cdots (1\pm f_X)(1\pm f_Y)\cdots. \label{eq:partial_probability}
\end{eqnarray}

\subsubsection{Expression of collision term}
The variation of the distribution $\Delta f$ through the microscopic process can be evaluated by
\begin{eqnarray}
 \Delta f_\phi \sim \sum_{\rm all \ processes}\Delta N_\phi \cdot \left[-\mathit{p}_\phi(\phi,A,B,\cdots \rightarrow X,Y,\cdots)+\mathit{p}_\phi(X,Y,\cdots \rightarrow \phi, A,B,\cdots)\right]
\end{eqnarray}
where $\phi$ is the focusing species, $\Delta N_\phi$ is a changing number of the quantum $\phi$ in the process $\phi,A,B,\cdots \leftrightarrow X,Y,\cdots$ ($\Delta N_\phi=1$ in this case), and $p_\phi$ is the partial transition probability for $\phi$ derived in (\ref{eq:partial_probability}).  Finally, the collision term can be evaluated as
\begin{eqnarray}
 \mathbf{C}[f_\phi]
  &\sim& \left.E_\phi\frac{\Delta f_\phi}{\Delta t}\right|_{\rm microscopic \ process} \\
  &=& -\frac{1}{2}\sum_{\rm all \ processes} \prod_{a\neq \phi} \left(\int \frac{d^3\vec{k}_a}{(2\pi)^3}\frac{1}{2E_a}\sum_{g_a}\right) \nonumber \\
  & & \qquad \times (2\pi)^4\delta^4(k_A+k_B+\cdots - k_X -k_Y-\cdots) \nonumber \\
  & & \qquad \times \Delta N_\phi\left[\left|\mathcal{M}(k_\phi,k_A,k_B,\cdots\rightarrow k_X,k_Y,\cdots)\right|^2 \right. \nonumber \\
  & & \qquad \qquad \qquad \qquad \times f_\phi f_Af_B\cdots (1\pm f_X)(1\pm f_Y)\cdots \nonumber \\
  & & \qquad \qquad \qquad -\left|\mathcal{M}(k_X,k_Y,\cdots \rightarrow k_\phi,k_A,k_B,\cdots )\right|^2 \nonumber \\
  & & \qquad \qquad \qquad \qquad \left.\times f_Xf_Y\cdots (1\pm f_\phi)(1\pm f_A)(1\pm f_B)\cdots\right]. \label{eq:Boltzmann_collision}
\end{eqnarray}
In the above derivation, we set $\Delta t =T$.  We derived the above result with the single-particle state for all species for simplicity.  In the case of including the $N$-duplicated species in $\phi, A, B, \cdots$ or $X,Y,\cdots$, one needs to multiply an extra factor $1/N!$ for the species.

Note that all the squared amplitudes in (\ref{eq:Boltzmann_collision}) must be regarded as the subtracted state in which the contribution of on-shell particles in the intermediate processes is subtracted in order to avoid the double-counting of the processes.  Such a situation will be faced in which the leading contributions of the amplitude consist of the loop diagrams or higher order of couplings, e.g., the baryogenesis scenario as we discuss later.

\subsection{Full and integrated Boltzmann equation}
Eqs. (\ref{eq:Boltzmann_geodesic}) and (\ref{eq:Boltzmann_collision}) lead the full Boltzmann equation for a species $\phi$ on the FLRW space-time as
\begin{eqnarray}
 \dot{f}_\phi-H\frac{|\vec{k}_\phi|^2}{E_\phi}\frac{\partial f_\phi}{\partial E_\phi}
  &=& -\frac{1}{2E_\phi}\sum_{\rm all \ processes} \prod_{a\neq \phi} \left(\int \frac{d^3\vec{k}_a}{(2\pi)^3}\frac{1}{2E_a}\sum_{g_a}\right) \nonumber \\
  & & \qquad \times (2\pi)^4\delta^4(k_A+k_B+\cdots - k_X -k_Y-\cdots) \nonumber \\
  & & \qquad \times \Delta N_\phi\left[\left|\mathcal{M}(k_\phi,k_A,k_B,\cdots\rightarrow k_X,k_Y,\cdots)\right|^2 \right. \nonumber \\
  & & \qquad \qquad \qquad \qquad \times f_\phi f_Af_B\cdots (1\pm f_X)(1\pm f_Y)\cdots \nonumber \\
  & & \qquad \qquad \qquad -\left|\mathcal{M}(k_X,k_Y,\cdots \rightarrow k_\phi,k_A,k_B,\cdots )\right|^2 \nonumber \\
  & & \qquad \qquad \qquad \qquad \left.\times f_Xf_Y\cdots (1\pm f_\phi)(1\pm f_A)(1\pm f_B)\cdots\right]. \label{eq:full_Boltzmann}
\end{eqnarray}

Eqs. (\ref{eq:full_Boltzmann}) for all species describe the detail evolution of the distribution functions, but they are not useful to solve because of a lot of variables.  To simplify the equations, the integrated Boltzmann equation is useful and convenient.  The momentum integral of the left hand side of (\ref{eq:full_Boltzmann}) leads
\begin{eqnarray}
 \int \frac{d^3\vec{k}_\phi}{(2\pi)^3} \sum_{g_\phi}\: ({\rm LHS \ of \ }
(\ref{eq:full_Boltzmann}) )
 &=& \dot{n}_\phi+3Hn_\phi \label{eq:integrated_Boltzmann_lhs}
\end{eqnarray}
where
\begin{equation}
 n_\phi\equiv \int \frac{d^3\vec{k}_\phi}{(2\pi)^3} \sum_{g_\phi}f_\phi
\end{equation}
is the number density of $\phi$.  Note in the integration (\ref{eq:integrated_Boltzmann_lhs}) that the variables $t$ and $|\vec{k}|=\sqrt{E^2-m^2}$ are independent and a property of the total derivative
\begin{equation}
 \int \frac{d^3\vec{k}}{(2\pi)^3}\:\frac{|\vec{k}|^2}{E}\frac{\partial}{\partial E}(\cdots) \:\:=\:\: \int \frac{d^3\vec{k}}{(2\pi)^3}\:\vec{k}\cdot\frac{\partial}{\partial \vec{k}}(\cdots) \:\:=\:\: -3\int \frac{d^3\vec{k}}{(2\pi)^3} (\cdots)
\end{equation}
is used.  As the result, the number of the dynamical variables are reduced from ($\#$species)$\times(\#t)\times(\#E)$ to ($\#$species)$\times (\#t)$, while the right hand side of the integrated (\ref{eq:full_Boltzmann}) still includes the $E$-dependent distribution functions.  An useful approximation is to apply the Maxwell-Boltzmann similarity distribution\footnote{The well-used approximation $f_a\sim\frac{n_a}{n_a^{\rm MB}} f_a^{\rm MB}$ can be justified in the case that species $a$ is in the kinetic equilibrium through interacting with the thermal bath.  See appendix \ref{sec:derivation_MB_approx} for detail.  The case in deviating from the kinetic equilibrium is discussed in section  \ref{sec:temperature_parameter}.
}
\begin{eqnarray}
 f_a, f_a^{\rm MB} \ll 1 \qquad {\rm and} \qquad f_a(t,E_a)\sim\frac{n_a(t)}{n_a^{\rm MB}(t)}f_a^{\rm MB}(t,E_a), \label{eq:MB_approx}
\end{eqnarray}
where
\begin{eqnarray}
 f_a^{\rm MB}=\exp\left[-\frac{E_a-\mu_a}{T_a}\right], \qquad n_a^{\rm MB}=\int \frac{d^3\vec{k}_\phi}{(2\pi)^3} \sum_{g_\phi}f_a^{\rm MB}
\end{eqnarray}
are the Maxwell-Boltzmann distribution and its number density, respectively.  Supposing the identical temperature for all species $T_\phi=T_A=\cdots\equiv T$, and substituting (\ref{eq:MB_approx}) into the right hand side of (\ref{eq:integrated_Boltzmann_lhs}) and assuming the chemical equilibrium
\begin{equation}
 \mu_\phi+\mu_A+\mu_B+\cdots=\mu_X+\mu_Y+\cdots
\end{equation}
where $\mu_a$ is the chemical potential of species $a$, one can obtain the integrated Boltzmann equation as
\begin{eqnarray}
 \dot{n}_\phi+3Hn_\phi
  &=& \int \frac{d^3\vec{k}_\phi}{(2\pi)^3} \sum_{g_\phi}\: ({\rm RHS \ of \ }
(\ref{eq:full_Boltzmann}) ) \nonumber \\
  &=& -\sum_{\rm all \ processes} \Delta N_\phi\left[n_\phi n_A n_B\cdots \times \langle R(\phi,A,B,\cdots \rightarrow X,Y,\cdots)\rangle \right. \nonumber \\
  & & \qquad \left. -n_\phi^{\rm MB}n_A^{\rm MB}n_B^{\rm MB}\cdots\frac{n_Xn_Y\cdots}{n_X^{\rm MB}n_Y^{\rm MB}\cdots} \times \langle R(\bar{\phi},\bar{A},\bar{B},\cdots\rightarrow \bar{X},\bar{Y},\cdots)\rangle\right] \nonumber \\ \label{eq:integrated_Boltzmann}
\end{eqnarray}
where the bar ``$\bar{\quad}$'' denotes its anti-particle state,
\begin{eqnarray}
 R(A,B,\cdots \rightarrow X,Y,\cdots)
  &\equiv& \frac{1}{2E_A2E_B\cdots}\int\frac{d^3k_X}{(2\pi)^3}\frac{d^3k_Y}{(2\pi)^3}\cdots\frac{1}{2E_X2E_Y\cdots} \nonumber \\
  & & \qquad \times \frac{\sum_{g_A,g_B,\cdots,g_X,g_Y,\cdots}|\mathcal{M}(k_A,k_B\cdots\rightarrow k_X,k_Y,\cdots)|^2}{\sum_{g_A,g_B,\cdots}} \nonumber \\
  & & \qquad \times (2\pi)^4\delta^4(k_A+k_B+\cdots - k_X-k_Y-\cdots) \label{eq:def_rate}
\end{eqnarray}
is a reaction rate integrated over the final state, and
\begin{eqnarray}
 \langle R(A,B,\cdots \rightarrow X,Y,\cdots)\rangle
  &\equiv& \frac{1}{n_A^{\rm MB}n_B^{\rm MB}\cdots}\int\frac{d^3k_A}{(2\pi)^3}\frac{d^3k_B}{(2\pi)^3}\cdots f_A^{\rm MB} f_B^{\rm MB} \cdots \nonumber \\
  & & \qquad \qquad \qquad \times \sum_{g_A,g_B,\cdots}R(A,B,\cdots\rightarrow X,Y,\cdots)
\end{eqnarray}
is the thermally averaged reaction rate by the Maxwell-Boltzmann distributions of the species appearing in the initial state.  We used the property of the CPT-invariance of the amplitude
\begin{equation}
 \mathcal{M}(X,Y,\cdots \rightarrow A,B,\cdots)=\mathcal{M}(\bar{A},\bar{B},\cdots\rightarrow \bar{X},\bar{Y},\cdots)
\end{equation}
to derive the last term in (\ref{eq:integrated_Boltzmann}).

Although eq. (\ref{eq:integrated_Boltzmann}) describing the evolution of number densities is obtained by integration of (\ref{eq:full_Boltzmann}) directly, in general, the other evolution equations of the statistical quantities
\begin{equation}
 Q_a(t) \equiv \int\frac{d^3\vec{k}_a}{(2\pi)^3}\sum_{g_a} f_a(\vec{k}_a)q_a(t,E_a)
\end{equation}
can also be derived through the same procedure with the corresponding coefficient $q_a(t,E_a)$, e.g., the energy density $Q=\rho$ for $q=E_a$, and the pressure $Q=P$ for $q=|\vec{k}|^2/3E_a$.  Such equations help to extract more detailed thermodynamic variables, e.g., to determine the independent temperatures for each species, as we will see in the next subsection.

\subsection{Temperature parameter} \label{sec:temperature_parameter}
The integrated Boltzmann equation (\ref{eq:integrated_Boltzmann}) is quite useful and can be applied to many situations.  However, it might not be suitable for some situations in which the kinetic equilibrium is highly violated because the formula is based on the approximation by the Maxwell-Boltzmann similarity distribution, which is justified by the kinetic equilibrium of the target particles with the thermal bath as discussed in appendix \ref{sec:derivation_MB_approx}.  Although to obtain the most appropriate solution is to solve the full Boltzmann equation, it takes a lot of costs to the calculation.  In this section, we introduce an alternative method based on the integrated Boltzmann equation.

Instead of using the similarity distribution (\ref{eq:MB_approx}), we introduce more generalized
similarity distribution by\footnote{The normalization factor $n_a(t)/n_a^{\rm neq}(t)$ can be regarded as a corresponding quantity to the {\it chemical potential parameter} $\tilde{\mu}_a(t)$:
\begin{eqnarray}
 \frac{n_a(t)}{n_a^{\rm neq}(t)}
  &=& \exp\left[\frac{\tilde{\mu}_a(t)-\mu_a}{T_a(t)}\right].
\end{eqnarray}
}
\footnote{In the case of the Bose-Einstein/Fermi-Dirac type distribution
\begin{equation}
 f_a(t,E_a)=\left[e^{(E_a-\tilde{\mu}_a(t))/T_a(t)}\mp 1\right]^{-1} \qquad ( \ -: \: {\rm boson}, \quad +: \: {\rm fermion}) \label{eq:BE-FD_like_distribution}
\end{equation}
where $T_a(t)$ and $\tilde{\mu}_a(t)$ are the temperature and the chemical potential parameters respectively, the temperature parameter can be represented as
\begin{equation}
 \tilde{T}_a(t) = \frac{1}{\rho_a(t)+P_a(t)}\int\frac{d^3\vec{k}_a}{(2\pi)^3}\frac{\vec{k}_a^2}{3}f_a(t,E_a)\left(1\pm f_a(t,E_a)\right) \qquad ( \ +: \: {\rm boson}, \quad -: \: {\rm fermion})
\end{equation}
with energy density $\rho_a$ and pressure $P_a$ evaluated by (\ref{eq:BE-FD_like_distribution}).  The above representation is consistent with (\ref{eq:temperature_parameter_relation}) in the nonrelativistic limit: $f_a\ll 1$, $E_a\sim m_a$, and  $\rho_a\sim m_a n_a\gg P_a$.  The chemical potential parameter can be obtained by
\begin{equation}
 \tilde{\mu}_a(t) = \frac{\rho_a(t)+p_a(t)-T_a(t)s_a(t)}{n_a(t)}
\end{equation}
where $s_a(t)$ is the entropy density defined by
\begin{equation}
 s_a(t)=\int\frac{d^3\vec{k}_a}{(2\pi)^3}\left[\pm(1\pm f_a)\ln(1\pm f_a)-f_a\ln f_a\right]. \qquad ( \ +: \: {\rm boson}, \quad -: \: {\rm fermion})
\end{equation}
}
\begin{equation}
 f_a(t,E_a) \sim \frac{n_a(t)}{n_a^{\rm neq}(t)}f_a^{\rm neq}(t,E_a), \qquad f_a^{\rm neq}(t,E_a)\equiv \exp\left[-\frac{E_a-\mu_a}{T_a(t)}\right].
\end{equation}
Here $n_a^{\rm neq}$ is the number density evaluated by the ``non-equilibrium'' Maxwell-Boltzmann distribution $f_a^{\rm neq}$ that is parametrized by the {\it temperature parameter} $T_a(t)$.  In general, the temperature parameter is independent of the thermal bath temperature $T(t)$.

Especially as the property of the Maxwell-Boltzmann distribution form, the temperature parameter can be expressed by the ratio of the pressure and the number density
\begin{equation}
 T_a \:\: = \:\: \frac{P_a^{\rm neq}}{n_a^{\rm neq}} \:\:=\:\: \frac{P_a}{n_a} \label{eq:temperature_parameter_relation}
\end{equation}
because of
\begin{eqnarray}
 P_a^{\rm neq} &=&\int\frac{d^3\vec{k}}{(2\pi)^3} \sum_{g_a}f_a^{\rm neq}\frac{|\vec{k}_a|^2}{3E_a} \:\:=\:\: \int\frac{d^3\vec{k}}{(2\pi)^3}\sum_{g_a}\frac{T_a}{3}\left(-\vec{p}_a\cdot \frac{\partial}{\partial\vec{p}_a}\right)f_a^{\rm neq} \nonumber \\
 &=& T_a\int\frac{d^3\vec{p}}{(2\pi)^3}\sum_{g_a}f_a^{\rm neq} \:\:=\:\: n_a^{\rm neq}T_a
\end{eqnarray}
and
\begin{equation}
 P_a= \int\frac{d^3\vec{k}}{(2\pi)^3}\sum_{g_a}f_a\frac{|\vec{k}_a|^2}{3E_a}=\frac{n_a}{n_a^{\rm neq}}\int\frac{d^3\vec{p}}{(2\pi)^3}\sum_{g_a}f_a^{\rm neq}\frac{|\vec{k}_a|^2}{3E_a}=\frac{n_a}{n_a^{\rm neq}}P_a^{\rm neq}.
\end{equation}
Therefore, the evolution equation for the temperature parameter can be derived from the pressure's one which can be constructed from the original full Boltzmann equation (\ref{eq:full_Boltzmann}) multiplied by $|\vec{k}_a|^2/3E_a$.  After the integration by the momentum, one can obtain the coupled equations for species $\phi$ as
\begin{eqnarray}
 & & \dot{n}_\phi+3Hn_\phi \nonumber \\
 & & \:\: = \:\: -\sum_{\rm all \ processes} \Delta N_\phi \left[ n_\phi n_A n_B\cdots \times \langle R(\phi,A,B,\cdots \rightarrow X,Y,\cdots)\rangle^{\rm neq}\right. \nonumber \\
 & & \qquad \qquad \qquad \qquad \left. -n_X n_Y\cdots \times \langle R(X,Y,\cdots \rightarrow \phi,A,B,\cdots)\rangle^{\rm neq}\right], \label{eq:integrated_Boltzmann_neq1} \\
 & & n_\phi \dot{T}_\phi+Hn_\phi \left(2T_\phi-\left<\frac{|\vec{k}_\phi|^4}{3E_\phi^3}\right>^{\rm neq}\right) \nonumber \\
 & & \:\: = \:\:  -\sum_{\rm all \ processes} \Delta N_\phi \left[ n_\phi n_A n_B\cdots \times \left<\left(\frac{|\vec{k}_\phi|^2}{3E_\phi}-T_\phi\right) R(\phi,A,B,\cdots \rightarrow X,Y,\cdots)\right>^{\rm neq}\right. \nonumber \\
 & & \qquad \qquad \qquad \qquad \qquad -n_X n_Y\cdots \times \left(\langle R_{T_\phi}(X,Y,\cdots\rightarrow \phi,A,B,\cdots)\rangle^{\rm neq}\right. \nonumber \\
 & & \quad \qquad \qquad \qquad \qquad \qquad \qquad \qquad \qquad \left. \left. -T_\phi \langle R(X,Y,\cdots\rightarrow \phi,A,B,\cdots)\rangle^{\rm neq}\right)\right], \label{eq:integrated_Boltzmann_neq2}
\end{eqnarray}
where $R$ is a rate defined in (\ref{eq:def_rate}) and
\begin{eqnarray}
 & & R_{T_\phi}(X,Y,\cdots\rightarrow\phi,A,B,\cdots) \nonumber \\
 & & \:\: = \:\: \frac{1}{2E_X2E_Y\cdots}\int\frac{d^3\vec{k}_\phi}{(2\pi)^3}\frac{d^3\vec{k}_A}{(2\pi)^3}\frac{d^3\vec{k}_B}{(2\pi)^3}\cdots \nonumber \\
  & & \qquad \times \frac{\sum_{g_\phi,g_A,g_B,\cdots,g_X,g_Y,\cdots}|\mathcal{M}(k_X,k_Y,\cdots\rightarrow k_\phi,k_A,k_B,\cdots)|^2}{\sum_{g_X,g_Y,\cdots}}\cdot \frac{|\vec{k}_\phi|^2}{3E_\phi} \nonumber \\
  & & \qquad \times (2\pi)^4\delta^4(k_\phi+k_A+k_B+\cdots-k_X-k_Y-\cdots)
\end{eqnarray}
is a ``temperature weighted'' rate, and
\begin{eqnarray}
 \left< \frac{|\vec{k}_\phi|^4}{3E_\phi^3}\right>^{\rm neq}
  &=& \frac{1}{n_\phi^{\rm neq}}\int\frac{d^3\vec{k}_\phi}{(2\pi)^3}\sum_{g_\phi} f_\phi^{\rm neq} \cdot \frac{|\vec{k}_\phi|^4}{3E_\phi^3}, \\
 \langle R(a,b, \cdots\rightarrow i,j,\cdots)\rangle^{\rm neq}
  &=& \frac{1}{n_a^{\rm neq}n_b^{\rm neq}\cdots}\int\frac{d^3\vec{k}_a}{(2\pi)^3}\frac{d^3\vec{k}_b}{(2\pi)^3}\cdots\sum_{g_a,g_b,\cdots} \nonumber \\
  & & \qquad \qquad \times f_a^{\rm neq}f_b^{\rm neq} \cdots R(a,b,\cdots\rightarrow i,j,\cdots),
\end{eqnarray}
are the thermally averaged quantities by the non-equilibrium distribution $f^{\rm neq}$ including only the initial species $a,b,\cdots$, not the final species $i,j,\cdots$.  Solving the coupled equations (\ref{eq:integrated_Boltzmann_neq1}) and (\ref{eq:integrated_Boltzmann_neq2}) for all the species can be expected to obtain more accurate results than the former integrated Boltzmann equation (\ref{eq:integrated_Boltzmann}).  Following the evolution in practice, the combined quantity
\begin{equation}
 y \:\: = \:\: \frac{m_\phi T_\phi}{s^{2/3}} \:\: \propto \:\: \frac{T_\phi}{T^2}
\end{equation}
instead of the solo $T_\phi$, where $s$ is the entropy density, is convenient for the non-relativistic $\phi$ because of the asymptotic behavior $T_\phi(t)\propto a(t)^{-2}\propto T(t)^2$ after freezing out.

\subsection{Limitation of the approach by the Boltzmann equation} \label{sec:BE_limitation}
Finally, we mention the limitation of the Boltzmann equation.  As seen in the derivation of the Boltzmann equation from the quantum field theory, the collision term is derived from the transition probability that is described by the $S$-matrix element.  In the usual method by the quantum field theory, the evaluated transition probability or the corresponding  $S$-matrix elements indicates the process from the infinite past to the infinite future, in both limits of which the distinct one-particle (or multi-particle) state can be well-defined.  In reality, however, the interaction process happens within a finite time and occurs continuously before the interacting state returns to the actual particle state.  If the evolution of the system is described by such a quasi-particle effect (such as involving a significant thermal corrective effect, a resonant effect in the elementary process, a quantum oscillation effect, etc.), one should consider a more accurate treatment of the system.

One of the methods is to consider the Kadanoff-Baym equation \cite{Baym:1961zz,Baym:1962sx} that describes the evolution of the two-point correlation functions with the self-energy effect through the closed time path formalism (or the Schwinger-Keldysh formalism) \cite{Chou:1984es}.  The calculated two-point functions can lead to the number density, the energy density, or so.  Although the formulation is more complicated than the Boltzmann equation, a more accurate treatment including the statistical quantum effects is possible.  The applications for the cosmological context can be seen in, e.g., \cite{Hamaguchi:2011jy,Binder:2018znk} for DM relic abundance, \cite{Riotto:1995hh,Buchmuller:2000nd,Garbrecht:2003mn,DeSimone:2007gkc,Cirigliano:2009yt,Beneke:2010dz} for various baryogenesis scenarios.  See also \cite{Berges:2004yj} for a pedagogical review of nonequilibrium quantum field theory and \cite{Garbrecht:2018mrp} for a recent review of baryogenesis scenarios based on the Kadanoff-Baym equation.

\section{Application to DM abundance} \label{sec:dark_matter}
One of the cosmological application of the Boltzmann equation is for the estimation of the DM abundance.  Because DM is stable, the main process changing the particle number is not decay/inverse-decay but the 2-2 annihilation/creation scatterings
\begin{equation}
 \chi, \bar{\chi} \leftrightarrow \psi, \bar{\psi}
\end{equation}
where $\chi$ is a DM and $\psi$ are a standard model particle.  Since the rate in the 2-2 scattering can be represented by the annihilation cross section as
\begin{equation}
 R(\chi, \bar{\chi} \rightarrow \psi, \bar{\psi})=\sigma v
\end{equation}
where $v$ is the M\o ller velocity\footnote{
The definition with the 4-momenta is given by
\begin{equation}
 v_{12} = \frac{\sqrt{(k_1\cdot k_2)^2-m_1^2m_2^2}}{k_1^0k_2^0},
\end{equation}
which can be identical to the relative velocity only in case of the parallel 3-momenta; $\vec{k}_1\cdot \vec{k}_2=\pm|\vec{k}_1||\vec{k}_2|.$} for the pair of the DM particles,
the dynamics can be solve as the annihilation cross section is given.  Assuming the symmetric DM $n_\chi=n_{\bar{\chi}}$ and the thermal distribution for the standard model particles $n_\psi=n_{\bar{\psi}}=n_\psi^{\rm MB}$, the Boltzmann equation (\ref{eq:integrated_Boltzmann}) for the DM leads a simple form
\begin{eqnarray}
 \dot{n}_\chi+3Hn_\chi=-\left(n_\chi^2-(n_\chi^{\rm MB})^2\right)\langle\sigma v\rangle. \label{eq:Boltzmann_eq_anninilation_process}
\end{eqnarray}
Instead of the particle number to follow its evolution by time, it is convenient to use the yield $Y_\chi \equiv n_\chi /s$ with a dynamical variable $x\equiv m_\chi /T$, where $s=\frac{2\pi^2}{45}h_{\rm eff}(T)T^3$ is the entropy density and $h_{\rm eff}(T)\sim 100$ for $T\gtrsim 100$ GeV is the effective degrees of freedom defined by the entropy density.  In the case of no creation/annihilation process, the yield $Y_\chi$ becomes a constant since the number and the entropy in the comoving volume is conserved.  With these variables, the Boltzmann equation (\ref{eq:Boltzmann_eq_anninilation_process}) can be represented as
\begin{equation}
 Y_\chi'=-(1+\delta_h)\frac{s\langle\sigma v\rangle}{xH}\left(Y_\chi^2-(Y_\chi^{\rm MB})^2\right) \label{eq:Boltzmann_yield}
\end{equation}
where we denote $'\equiv d/dx,$ and
\begin{equation}
 \delta_h\equiv \frac{T}{3h_{\rm eff}}\frac{d h_{\rm eff}}{d T}.
\end{equation}
Since the adiabatic parameter $\delta_h$ tends to be negligible in the almost era of the thermal history\footnote{
If the DM mass scale is around $\mathcal{O}(10)$ GeV, the freeze-out occurs around the QCD transition scale $T\sim \mathcal{O}(100)$ MeV, in which $|\delta_h| \sim \mathcal{O}(1)$.  Thus, there is a few percent level contribution from the adiabatic parameter $\delta_h$ even in the WIMP model.  See Refs. \cite{Srednicki:1988ce,Gondolo:1990dk,Hindmarsh:2005ix,Drees:2015exa,Laine:2015kra,Borsanyi:2016ksw,Saikawa:2018rcs,Saikawa:2020swg} for the determination of that parameter in detail.},
we set $\delta_h=0$ in the later discussion for simplicity.  Moreover, we denote
\begin{equation}
 Y_\chi^{\rm MB} \:\:\equiv\:\: \frac{n_\chi^{\rm MB}}{s} \:\:\sim\:\: \frac{g_\chi}{h_{\rm eff}}\frac{45}{2^{5/2}\pi^{7/2}}x^{3/2}e^{-x} \qquad (x\gg 1),
\end{equation}
where $g_\chi$ is the degrees of freedom for the DM particle.

\subsection{Relic abundance in freeze-out}
As a simple and reasonable setup, we assume that the DM particles $\chi$ are in thermal equilibrium initially.  Then, the dynamics described by (\ref{eq:Boltzmann_yield}) can be explained as follow.  At first, the system is in the thermal equilibrium due to the stronger scattering effect than the spatial expansion\footnote{If the interaction rate becomes lower than the Hubble rate at the relativistic regime $x\lesssim1$, the abundance freezes out with the massless abundance (hot relic):
\begin{equation}
 Y_\infty \:\:\sim \:\: Y_{\rm hot}=\frac{45\zeta(3)}{2\pi^4}\frac{g_\chi}{h_{\rm eff(T_f)}}\times \left\{ \begin{array}{cc} 1 & ({\rm boson}) \\ 3/4 & ({\rm fermion}) \end{array} \right.
\end{equation}
where $\zeta(3)=1.202\cdots$.}, but the yield has a small deviation from the thermal value due to the expansion effect as
\begin{equation}
 Y_\chi\sim Y_\chi^{\rm MB}+\Delta(x), \qquad \Delta(x)=\frac{xH}{s\langle \sigma v\rangle}\frac{-Y_\chi'}{Y_\chi^{\rm MB}+Y}\sim \frac{xH}{2s\langle \sigma v\rangle}\ll Y_\chi^{\rm MB}
\end{equation}
as long as $n^{\rm MB}\langle\sigma v\rangle \gg xH$.  The deviation $\Delta$ continues growing in later time, and finally the evolution of the yield freezes out because the expansion rate exceeds the scattering rate.  The freeze-out occurs when $\Delta(x_f)=cY^{\rm MB}(x_f)$, $c\sim \mathcal{O}(1)$.  The freeze-out time $x=x_f$ and the final abundance $Y_\infty=Y(x=\infty)$ can be estimated by \cite{Kolb:1990vq,Cline:2013gha}
\begin{eqnarray}
 x_f
  &=& \ln \left[c(c+2)\frac{\sqrt{90}}{(2\pi)^3}\frac{g_\chi}{\sqrt{g_{\rm eff}(T_f)}}m_\chi M_{\rm pl}\sigma_n\right]-\left(n+\frac{1}{2}\right)\ln x_f, \label{eq:xf} \\
  &=& \ln \left[c(c+2)\frac{\sqrt{90}}{(2\pi)^3}\frac{g_\chi}{\sqrt{g_{\rm eff}(T_f)}}m_\chi M_{\rm pl}\sigma_n\right] \nonumber \\
  & & -\left(n+\frac{1}{2}\right)\ln \left(\ln \left[c(c+2)\frac{\sqrt{90}}{(2\pi)^3}\frac{g_\chi}{\sqrt{g_{\rm eff}(T_f)}}m_\chi M_{\rm pl}\sigma_n\right]\right)+\cdots, \\
 Y_\infty
  &=& (n+1)\sqrt{\frac{45}{\pi}}\frac{g_\chi}{\sqrt{g_{\rm eff}(T_f)}}\frac{x_f^{n+1}}{M_{\rm pl}m_\chi\sigma_n} \label{eq:Y_inf}
\end{eqnarray}
where $M_{\rm pl}=1.22\times 10^{19}$ GeV is the Planck mass, $g_{\rm eff}(T)$ is the effective degrees of freedom defined by the energy density $\rho=\frac{\pi^2}{30}g_{\rm eff}(T)T^4$, and $T_f=m_\chi/x_f$ is the freeze-out temperature.  In the derivation of the analytic results (\ref{eq:xf}) and (\ref{eq:Y_inf}), the temperature dependence of the cross section is approximated by the most dominant part as
\begin{equation}
 \langle \sigma v\rangle=\sigma_n x^{-n},
\end{equation}
where $\sigma_n$ is a constant\footnote{See also Appendix \ref{sec:formulae_TA} for the actual analysis of the thermally averaged cross section.}.  Especially, $n=0$ and $1$ correspond to $s$-wave and $p$-wave scattering, respectively.  Although a numerical factor $c$ still has uncertainty, choosing $c(c+2)=n+1$ leads to better analysis for the final abundance $Y_\infty$ within $5\%$ accuracy for $x_f\gtrsim 3$ \cite{Kolb:1990vq}.

\begin{figure}[t]
 \centering
 \includegraphics[keepaspectratio, scale=1]{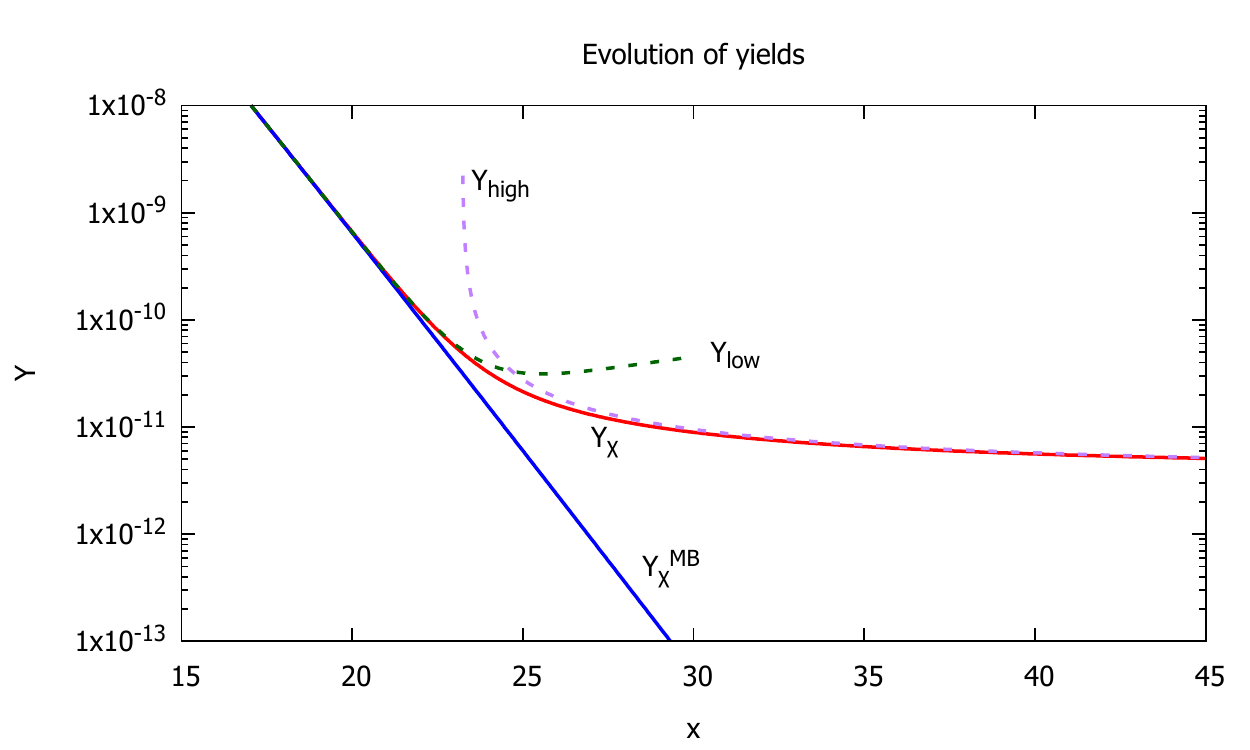}
 \caption{The numerical plots of the evolution for each yield with parameters $m_\chi=120$ GeV, $n=1$, $\sigma_n=\frac{\alpha_W^2}{m_\chi^2}$, $\alpha_W=\frac{1}{30}$, $g_\chi=2$, $h_{\rm eff}=g_{\rm eff}=90$.  The red and the blue lines show the actual evolution of $Y_\chi$ and the thermal yield $Y_\chi^{\rm MB}$, respectively.  The dashed lines of green and purple show the approximated solutions $Y_{\rm low}\equiv Y_\chi^{\rm MB}+\Delta$ and $Y_{\rm high}\equiv Y_\infty\left[1-\frac{Y_\infty}{n+1}\frac{s\langle\sigma v\rangle}{H}\right]^{-1}$, respectively.}
 \label{fig:darkmatter}
\end{figure}

As an example, let us consider a WIMP model.  Choosing the parameters as $m_\chi=120$ GeV, $n=1$, $\sigma_n=\frac{\alpha_W^2}{m_\chi^2}$, $\alpha_W=\frac{1}{30}$, $g_\chi=2$, $h_{\rm eff}=g_{\rm eff}=90$, one can obtain the analytic results
\begin{equation}
 x_f=23.2, \qquad Y_\infty=3.81\times 10^{-12}. \label{eq:ex_wimp}
\end{equation}
The actual evolution is depicted in Figure \ref{fig:darkmatter}.

Finally, we need to mention the validity of the approximated results (\ref{eq:xf}) and (\ref{eq:Y_inf}).  Their behaviors can deviate easily if the master equation (\ref{eq:Boltzmann_eq_anninilation_process}) includes the significant extra processes by other species or the singular behavior of the cross sections.   Especially it is known some exceptional cases; (i) mutual annihilations of multiple species (\textit{coannihilations}), (ii) annihilations into heaver states (\textit{forbidden channels}), (iii) annihilations near a pole in the cross section \cite{Griest:1990kh}, and (iv) simultaneous chemical and kinetic decoupling  (\textit{coscattering}) \cite{DAgnolo:2017dbv}.  In these cases, the analysis should be performed more carefully.  See \cite{Griest:1990kh,DAgnolo:2015ujb,DAgnolo:2017dbv,Kim:2019udq,Arkani-Hamed:2006wnf,Tulin:2012uq,Ibarra:2015nca} as their example cases, and also \cite{Bringmann:2006mu,Bringmann:2009vf,Binder:2017rgn} as examples of the evaluation with the temperature parameter.  As comparisons with their results, recent papers \cite{Ala-Mattinen:2019mpa,Ala-Mattinen:2022nuj} show results by the full Boltzmann equation without temperature parameter.

\subsection{Constraint on relic abundance}
The relic abundance for the stable particles through the freeze-out of their annihilation processes, as similar to $\chi$ particles discussed in the above, are restricted by the cosmological observation results.  
An useful parameter relating to the relic abundance is the density parameter defined by
\begin{equation}
 \Omega_\chi\equiv \frac{\rho_\chi}{\frac{3M_{\rm pl}^2}{8\pi}H^2}\sim\frac{16\pi^3}{135}\frac{h_{\rm eff}(T)T^3}{M_{\rm pl}^2H^2}\cdot m_\chi Y_\chi. \label{eq:density_parameter_chi}
\end{equation}
Since the yield maintains the constant after the freeze-out unless the additional entropy production occurs in the later era, one can estimate the present density parameter of $\chi$ with the present values.   The current observation through the the Cosmic Microwave Background \cite{Planck:2018nkj} provides $T_0=2.726$ K, $h_{\rm eff}(T_0)=3.91$, $H_0=100h^2$ km/s/Mpc, $h=0.677$, therefore one can estimate to
\begin{equation}
 \Omega_{\chi,{\rm now}}\sim\frac{m_\chi Y_\chi}{3.64h^2\times 10^{-9} \ {\rm GeV}}. \label{eq:density_parameter_chi_present}
\end{equation}
Because the present density parameter for the cold matter component is observed as $\Omega_c h^2=0.119$ and it must be larger than the $\chi$'s component, one can obtain a bound as
\begin{equation}
 m_\chi Y_\chi < 4.36\times 10^{-10} \ {\rm GeV}.
\end{equation}
The set of parameters shown in (\ref{eq:ex_wimp}) is seemingly suitable for the above constraint with a bit of the modification.  However, that would fail by taking into account the direct detections of the DM that focuses on the process of $\chi, \psi \leftrightarrow \chi, \psi$, where $\psi$ is a standard model particle.  If the annihilation process occurs through a similar interaction to the electroweak gauge interaction, the cross section for $\chi$-$\psi$ elastic scattering also relates to the same gauge interaction.  One can estimate $\sigma_{\chi\psi\rightarrow \chi\psi} \sim G_F^2 m_\chi^2 \sim 10^{-36} \ {\rm cm}^2$, but it is already excluded by the direct detection \cite{Roszkowski:2017nbc}.

\subsection{Relic abundance in freeze-in}
The discussion and the result in the previous subsections are based on the freeze-out scenario in which the DM particles are in thermal equilibrium initially.  However, it is not satisfied if the interaction between the DM particles and the thermal bath is too small, so-called FIMP (feebly interacting massive particle) scenario \cite{Hall:2009bx,Bernal:2017kxu}.  In this situation, the yield of DM evolves from zero through the thermal production from the thermal bath.  Although the DM never reaches the thermal equilibrium, the yield freezes in with a non-thermal yield at last.

We discuss here the relic abundance by the freeze-in scenario in two cases of the pair-creation of DM by (1) scattering from thermal scattering and (2) decay from a heavier particle.

\subsubsection{Pair-creation by scattering}
In the case that DM-pair ($\chi\bar{\chi}$) is produced by thermal pair particles ($\psi\bar{\psi}$), the Boltzmann equation is given by (\ref{eq:Boltzmann_yield}) as
\begin{eqnarray}
 Y_\chi'
  &\sim& \frac{s\langle\sigma v\rangle}{xH}(Y_\chi^{\rm MB})^2,
\end{eqnarray}
where we approximated $Y_\chi\ll Y_\chi^{\rm MB}$ and the adiabatic degrees $\delta_h\sim 0$ until the freezing-in.  For simplicity, we consider the simple interaction described by
\begin{equation}
 \mathcal{L}_{\rm int} \:\:=\:\: \lambda (\chi^\dagger \chi)(\psi^\dagger\psi).
\end{equation}
where $\chi$ is the bosonic DM, $\psi_i$ labeled $i$ are the massless bosons in the thermal bath, and $y$ is a coupling constant. The thermally averaged cross section is given by
\begin{eqnarray}
 \langle\sigma v\rangle
  &=& \frac{g_\chi^2g_\psi^2}{(n_\chi^{\rm MB})^2}\frac{\lambda^2}{(2\pi)^5} \: T\int_{4m_\chi^2}^\infty ds \: \sqrt{s-4m_\chi^2} \: K_1(\sqrt{s}/T).
\end{eqnarray}
where $g_\chi$ and $g_\psi$ are the degrees of freedom for each species.  Therefore, one can estimate the final yield at $x=m_\chi/T= \infty$ as
\begin{eqnarray}
 Y_\chi(\infty)
  &\sim& \int_0^\infty dx \: \frac{s\langle\sigma v\rangle}{xH}(Y_\chi^{\rm MB})^2 \\
  &=& \frac{3\pi^2}{128}\cdot g_\chi^2g_\psi^2\frac{\lambda^2}{(2\pi)^5}\cdot\frac{m_\chi^4}{H(T=m_\chi) \: s(T=m_\chi)}. \label{eq:final_Y_freze-in1}
\end{eqnarray}
This result implies that the yield freezing occurs around the earlier stage $x\sim \mathcal{O}(1)$ because (\ref{eq:final_Y_freze-in1}) can be regarded as $Y_\chi(\infty)\sim \left.\frac{n_\chi^{\rm MB}\langle\sigma v\rangle}{H}Y_\chi^{\rm MB}\right|_{x\sim 1}$.

Applying the obtained relic abundance (\ref{eq:final_Y_freze-in1}) to the relation of the present density parameter (\ref{eq:density_parameter_chi_present}), one can obtain the required strength of the coupling as
\begin{equation}
 \lambda \:\: = \:\: 1.0\times 10^{-12} \cdot \frac{1}{g_\chi g_\psi}\cdot \left(\frac{g_{\rm eff}(T=m_\chi)}{100}\right)^{1/4} \: \left(\frac{h_{\rm eff}(T=m_\chi)}{100}\right)^{1/2}\cdot\left(\frac{\Omega_{\chi,{\rm now}}h^2}{0.119}\right)^{1/2}.
\end{equation}

\subsubsection{Pair-creation by decay}
The other possible freeze-in scenario is due to the pair production from a heavier particle: $\sigma \rightarrow \chi \bar{\chi}$ \cite{Hall:2009bx,Chu:2011be}.  The original Boltzmann equation for DM is given by
\begin{eqnarray}
 \frac{dY_\chi}{dx_\sigma}
  &=& (1+\delta_h)\frac{\Gamma_{\sigma\rightarrow \chi\bar{\chi}}}{x_\sigma H}\frac{K_1(x_\sigma)}{K_2(x_\sigma)}\left(Y_\sigma-\left(\frac{Y_\chi}{Y_\chi^{\rm MB}}\right)^2 Y_\sigma^{\rm MB}\right) \\
  &\sim& \frac{\Gamma_{\sigma\rightarrow \chi\bar{\chi}}}{x_\sigma H}\frac{K_1(x_\sigma)}{K_2(x_\sigma)}Y_\sigma^{\rm MB}
\end{eqnarray}
where $\Gamma_{\sigma\rightarrow \chi\bar{\chi}}$ is a decay constant and $x_\sigma \equiv m_\sigma/T$.  We also approximated $Y_\sigma\sim Y_\sigma^{\rm MB}$, $Y_\chi\ll Y_\chi^{\rm MB}$, and $\delta_h\sim 0$ in the second line.  Therefore, the final yield can be estimated as
\begin{eqnarray}
 Y_\chi(\infty)
  &\sim& \int_0^\infty dx_\sigma\frac{\Gamma_{\sigma\rightarrow \chi\bar{\chi}}}{x_\sigma H}\frac{K_1(x_\sigma)}{K_2(x_\sigma)}Y_\sigma^{\rm MB}\\
  &=& \frac{3g_\sigma}{4\pi}\cdot \frac{\Gamma_{\sigma\rightarrow \chi\bar{\chi}}}{H(T=m_\sigma)}\frac{m_\sigma^3}{s(T=m_\sigma)}.
\end{eqnarray}
where $g_\sigma$ is the degrees of freedom for $\sigma$.  If the decay constant can be represented by the coupling constant $y$ as
\begin{equation}
 \Gamma_{\sigma\rightarrow \chi\bar{\chi}} \:\: = \:\: g_\chi \cdot \frac{y^2}{8\pi}m_\sigma,
\end{equation}
the required magnitude of the coupling with the relation formula to the density parameter (\ref{eq:density_parameter_chi_present}) can be estimated as
\begin{equation}
 y^2 \:\: \sim \:\: 2.7\times 10^{-24} \cdot \frac{m_\sigma}{m_\chi}\cdot \frac{1}{g_\sigma g_\chi}\cdot \left(\frac{g_{\rm eff}(T=m_\sigma)}{100}\right)^{1/2} \: \frac{h_{\rm eff}(T=m_\sigma)}{100}\cdot\frac{\Omega_{\chi,{\rm now}}h^2}{0.119}.
\end{equation}

\section{Application to baryogenesis}
The other popular application of the Boltzmann equation in cosmology is the baryogenesis scenario that describes the dynamical evolution of the baryon number in the Universe from zero at the beginning to the non-zero at present.  The present abundance of the baryons can be estimated from (\ref{eq:density_parameter_chi_present}).  Replacing $\chi$'s mass $m_\chi$ into the nucleon mass $m_N=939$ MeV and using the present density parameter for the baryon $\Omega_b h^2=0.0224$ \cite{Planck:2018nkj}, one can obtain
\begin{equation}
 Y_{B,{\rm now}}=8.69\times 10^{-11}. \label{eq:baryon_bound}
\end{equation}

There are three conditions suggested  by A.~D.~Sakharov \cite{Sakharov:1967dj} in order to develop the baryon abundance from $Y_B=0$ to non-zero: (1) baryon number ($B$) violation, (2) $C$ and $CP$ violation, (3) non-equilibrium condition.  Their brief reasons are as follows.  The $B$ violation is trivial by definition.  If the baryon number violating processes conserve $C$ or $CP$, their anti-particle processes happen with the same rate.  As the result, the net baryon number is always zero.  Even if the processes violate the baryon number, $C$ and $CP$, the thermal equilibrium reduces the baryon asymmetry due to their inverse processes.  Especially,  the Boltzmann equation provides a powerful tool to quantify the third condition.

To see how to construct the Boltzmann equations for the baryogenesis, let us consider with a toy model\footnote{Replacing $X$, $\psi$, $\phi$ into the right-handed neutrino, left-handed neutrino, Higgs doublet in the standard model, respectively, one can obtain the type-I seesaw model that can realize the well-known leptogenesis scenario \cite{Fukugita:1986hr}.  However, the correspondence is incomplete: the type-I seesaw model includes the gauge interactions that induces $\Delta L=1$ scattering process.  See e.g., Ref.\cite{Davidson:2008bu} for a review of the leptogenesis scenario and \cite{Anisimov:2010dk,Garbrecht:2018mrp} for the treatment by the Kadanoff-Baym equation.} as shown in Table \ref{tab:matter_contents}.  The model includes Majorana-type of chiral fermions $X_a$ for $a=1,\cdots, N_X$, baryonic chiral fermions $\psi_i$ for $i=1,\cdots,N_\psi$ with the common baryon number $b$, and a non-baryonic complex scalar $\phi$.  For simplicity, the baryonic fermions $\psi_i$ and the scalar $\phi$ are massless and they are always in the thermal equilibrium.  Because $X_a$ are the Majorana fermion, $X_a$ and $\bar{X}_a$ can be identified.  Thus, once we set the fundamental interaction to provide a decay/inverse-decay processes $X_a\leftrightarrow \psi_i, \phi$, their anti-particle processes $X_a\leftrightarrow \bar{\psi}_i,\bar{\phi}$ also exist.  These 3-body interactions also induce the $B$-violating 2-2 scatterings exchanging  $X_a$ fermions.

\begin{table}[t]
 \centering
 \begin{tabular}{c|c|c}
  \hline
  Species & Particle statistic & \#$B$ \\
  \hline \hline
  $X_a$ & Chiral fermion (Majorana) & $-$ \\
  \hline
  $\psi_i$ & Chiral fermion & $b$ \\
  \hline
  $\phi$ & Complex scalar & $0$ \\
  \hline
 \end{tabular}
 \hspace{5em}
 \begin{tabular}{c|c}
  \hline
  Process & $\Delta B$ \\
  \hline \hline
  $X_a\rightarrow \psi_i,\phi$ & $b$ \\
  \hline
  $X_a\rightarrow\bar{\psi}_i,\bar{\phi}$ & $-b$ \\
  \hline
  $\psi_i,\psi_j\rightarrow\bar{\phi},\bar{\phi}$ & $-2b$ \\
  \hline
  $\psi_i,\phi \rightarrow\bar{\psi}_j,\bar{\phi}$ & $-2b$ \\
  \hline
 \end{tabular}
 \caption{{\it Left}: the matter contents and their baryon number.  All the anti-particles have the opposite sign of the baryon number.  {\it Right}: Possible processes up to the 4-body $B$-violating interactions and their variation of the baryon number.  The bar ($\bar{\;\;}$) on each species denotes the anti-particle.  In addition to the shown processes here, their inverse processes are also possible.  Although there are elastic scatterings $X_a,\psi_i,\rightarrow X_b,\psi_j$, $X_a,\phi\rightarrow X_b,\phi$, and $\psi_i,\phi\rightarrow \psi_j,\phi$, we omited them because they do not change the baryon number.}
 \label{tab:matter_contents}
\end{table}

\subsection{Mean net baryon number}
At first, we consider only the decay processes for simplicity.  This situation is realized when $X_a$ particles start to decay after the scattering processes freeze out.  Here we define the mean net baryon number by
\begin{eqnarray}
 \epsilon_a
  &=& \sum_f\Delta B_f\left( r_{X_a\rightarrow f}- r_{\bar{X}_a\rightarrow \bar{f}}\right)
\end{eqnarray}
where the summation runs for all decay processes, $\Delta B_f$ and $r_{X_a\rightarrow f}$ are the generated baryon number through the process of $X_a\rightarrow f$ and its branching ratio, respectively.  The physical meaning of the mean net baryon number $\epsilon_a$ is an average of the produced baryon number by a single quantum of $X_a$.  In the case of our toy model, this quantity can be represented as
\begin{eqnarray}
 \epsilon_a
  &=& b\sum_i\left(r_{X_a\rightarrow \psi_i,\phi}-r_{X_a\rightarrow \bar{\psi}_i,\bar{\phi}}\right) \label{eq:mean_net_baryon}
\end{eqnarray}
This result reflect the requirements of $B$-violation and $C$, $CP$ violation.  If the decay processes are $B$-conserving $b=0$ or $C$, $CP$ conserving processes $r_{X_a\rightarrow \psi_i,\phi}=r_{X_a\rightarrow \bar{\psi}_i,\bar{\phi}}$, the mean net baryon number is vanished.

Supposing that only a single flavour $X_1$ survives and all of $X_1$ particles decay into the baryonic fermions $\psi_i$, the generated baryon number can be estimated by $Y_B\sim \epsilon_1Y_{X_1}$.  Especially, the baryon abundance can be maximized if $X_1$ particles are the hot relic $Y_{X_1}\sim Y_{\rm hot}\sim\frac{45}{2\pi^4}\frac{g_{X_1}}{h_{\rm eff}(T_f)}$:
\begin{equation}
 Y_B\sim \frac{45}{2\pi^4}\cdot \frac{\epsilon_1 g_{X_1}}{h_{\rm eff}(T_f)}, \label{eq:asym_sol_YB}
\end{equation}
where $g_{X_1}=2$ is the degrees of freedom of the Majorana-type fermion $X_1$.

\subsection{Boltzmann equations in baryogenesis scenario} \label{sec:boltzmann_eq_baryogenesis}
Although we considered quite simplified situation in the previous subsection, in reality, the situation is more complicated since the system includes the dynamical decay/inverse decay and scattering processes.  In order to quantify the actual evolution of the baryon abundance including the scattering effects, we need to construct the Boltzmann equations in this system and solve them.

For simplicity, we suppose again that only a single flavour $X_1$ affects to the evolution of the net baryon number.  Using the definition of the net baryon density $n_B= b\sum_i\left(n_{\psi_i}-n_{\bar{\psi}_i}\right)$, the evolution of the system including the processes in Table \ref{tab:matter_contents} are  described by\footnote{See appendix \ref{sec:formulae_TA} for the treatment of the thermally averaged quantities.  And also see Appendix \ref{sec:derivation_of_boltzmann_eq} for the detail of the derivation of the equations, especially, the treatment of the real intermediate state (RIS) to avoid the double-counting.}
\begin{eqnarray}
 \dot{n}_{X_1}+3Hn_{X_1}
  &=& -\left<\frac{M_{X_1}}{E_{X_1}}\right>\Gamma_{X_1}(n_{X_1}-n_{X_1}^{\rm MB})+\cdots, \label{eq:EOM_X}\\ \nonumber \\
 \dot{n}_B+3Hn_B
  &=& \epsilon_1\left<\frac{M_{X_1}}{E_{X_1}}\right>\Gamma_{X_1}\left(n_{X_1}- n_{X_1}^{\rm MB}\right) - 2\Gamma_S \: n_B +\cdots, \label{eq:EOM_net_baryon_1}
\end{eqnarray}
where $M_{X_1}$ ($E_{X_1}$) is the mass (energy) of $X_1$, $\Gamma_{X_1}\sim\sum_i\left(\Gamma_{X_1\rightarrow \psi_i\phi}+\Gamma_{X_1\rightarrow\bar{\psi}_i,\bar{\phi}}\right)$ is the total width of $X_1$ and
\begin{equation}
 \epsilon_1\equiv b\cdot \frac{\sum_i\left(\Gamma_{X_1\rightarrow \psi_i\phi}-\Gamma_{X_1\rightarrow\bar{\psi}_i,\bar{\phi}}\right)}{\Gamma_{X_1}} \label{eq:mean_net_baryon2}
\end{equation}
is the mean net baryon number corresponding to (\ref{eq:mean_net_baryon}), and
\begin{eqnarray}
 \Gamma_S
  &=& n_\phi^{\rm MB}\langle\sigma_{\psi\phi\rightarrow\bar{\psi}\bar{\phi}}v\rangle+n_\psi^{\rm MB}\langle\sigma_{\psi\psi\rightarrow\bar{\phi}\bar{\phi}}v\rangle
\end{eqnarray}
is the reaction rates through the B-violating scatterings up to the tree level.  The omitted parts ``$\cdots$'' denote the sub-leading processes in terms of the order of couplings.  Eq. (\ref{eq:EOM_X}) describes the dissipation of $X_a$, and it converts to the baryon with the rate $\epsilon_a$ and flows into the baryon sector.  However, the produced baryons also wash themselves out through the $B$-violating scattering processes due to the last term in (\ref{eq:EOM_net_baryon_1}).  Therefore, the smaller $B$-violating scattering effect is favored for remaining the more net baryons as long as $X_a$ can be thermalized enough at the initial.

To solve the equations of motion (\ref{eq:EOM_X}) and (\ref{eq:EOM_net_baryon_1}), it is convenient to use the yields $Y_{X_1}=n_{X_1}/s, Y_B=n_B/s$ and the variable $x=M_{X_1}/T$ as
\begin{eqnarray}
 Y_{X_1}' &=& -\gamma_D (Y_{X_1}-Y_{X_1}^{\rm MB}), \label{eq:eom_X_Y}\\
 Y_B' &=& \epsilon_1 \gamma_D (Y_{X_1}-Y_{X_1}^{\rm MB})-2\gamma_SY_B, \label{eq:eom_B_Y}
\end{eqnarray}
where
\begin{eqnarray}
 Y_{X_1}^{\rm MB}
  &=& \frac{n_{X_1}^{\rm MB}}{s} \:\: = \:\: \frac{45}{4\pi^4}\frac{g_{X_1}}{h_{\rm eff}}x^2K_2(x), \\
 \gamma_D &=& \frac{1}{x}\cdot \frac{\Gamma_{X_1}}{H(T)}\left<\frac{m_{X_1}}{E_{X_1}}\right> \:\: = \:\: \sqrt{\frac{45}{4\pi^3g_{\rm eff}}} \: \frac{M_{\rm pl}}{M_{X_1}}\cdot\frac{K_1(x)}{K_2(x)}\frac{\Gamma_{X_1}}{T}, \\
 \gamma_S &=& \frac{1}{x}\cdot \frac{\Gamma_S}{H(T)} \:\: = \:\: \sqrt{\frac{45}{4\pi^3g_{\rm eff}}} \: \frac{M_{\rm pl}}{M_{X_1}}\cdot\frac{\Gamma_S(T)}{T},
\end{eqnarray}
with the $n$-th order of the modified Bessel function $K_n(x)$.  We assumed the adiabatic evolution of the relativistic degrees $h_{\rm eff}'/h_{\rm eff}\sim 0$ to obtain (\ref{eq:eom_X_Y}) and (\ref{eq:eom_B_Y}).  The dimensionless parameters $\gamma_{D,S}$ are the reaction rates normalized by the Hubble parameter.
In general, $\gamma_D$ is proportional to $x^2$ ($x^1$) at the limit of $x\ll1$ ($x\gg 1$), whereas the behavior of $\gamma_S$ depends on the detail of the interaction as we will see its concrete form with an example model later.

Eqs.(\ref{eq:eom_X_Y}) and (\ref{eq:eom_B_Y}) can provide the analytic form of $Y_B$ as
\begin{eqnarray}
 Y_B(\infty)
  &=& -\epsilon_1\int_0^\infty dx \: Y_{X_1}'(x)\exp\left[-2\int_x^\infty dx'\:\gamma_S(x')\right]
\end{eqnarray}
Especially in the weakly scattering case, $\int_0^\infty dx \:\gamma_S\lesssim 1$, one can approximate the above result as
\begin{eqnarray}
 Y_B(\infty)
  &\sim& -\epsilon_1\int_0^\infty dx \: Y_{X_1}'(x) \:\: = \:\: \epsilon_1Y_{\rm hot}. \label{eq:YB_inf_approx1}
\end{eqnarray}
The physical interpretation is that the whole $X_1$ particles existing from the beginning can convert to the net baryons without any wash-out process in this case.  Hence the approximated result does not depend on the detail of the decay process $\gamma_D$.
The result (\ref{eq:YB_inf_approx1}) is consistent with the former estimation in (\ref{eq:asym_sol_YB}).
On the other hand, the strongly scattering case causes the wash-out process significantly, and thus the final net baryon abundance is strongly suppressed from the result of (\ref{eq:YB_inf_approx1}).

\begin{figure}[p]
 \centering
 \includegraphics[keepaspectratio, scale=1.0]{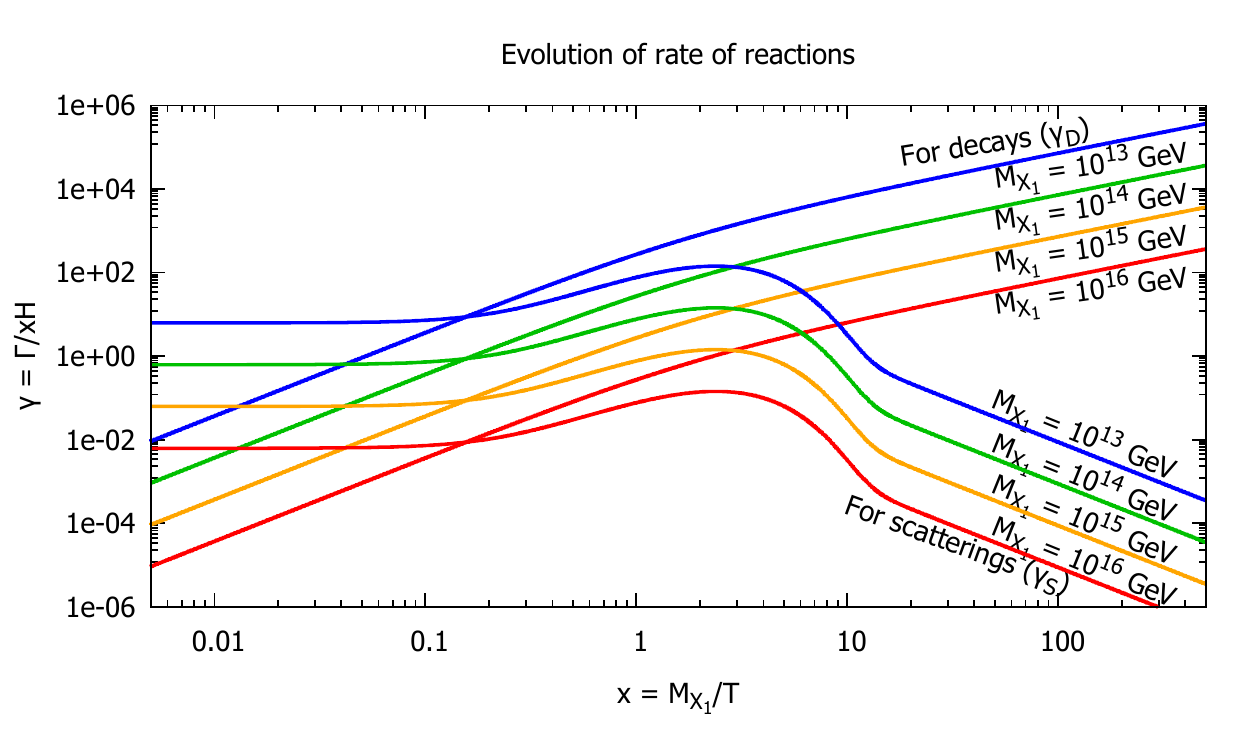}
 \includegraphics[keepaspectratio, scale=1.0]{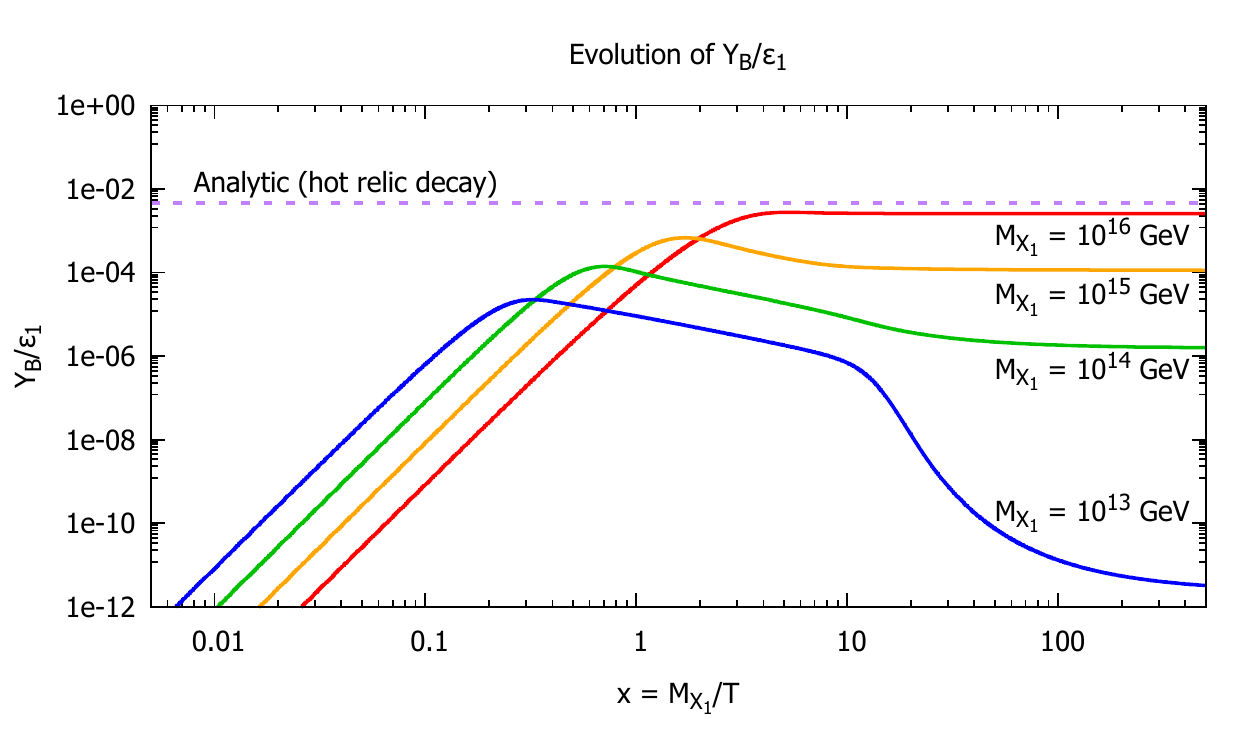}
 \caption{The numerical plots of the evolution of the interaction rates (upper) and $Y_B/\epsilon_1$ (lower) for each mass of $X$.  The numerical parameters are chosen as $g_{X_1}=2$, $g_{\psi_i}=g_\phi=1$, $N_\psi=3$, $\tilde{\alpha}=0.01$, $h_{\rm eff}=g_{\rm eff}=100$, and assumed the thermal distribution for $X_1$ and $Y_B=0$ at the initial.  The solid lines in red, yellow, green, and blue correspond to $M_{X_1}=10^{16},10^{15},10^{14}$, and $10^{13}$ GeV, respectively.  In the upper figure, ``For decays'' and ``For scatterings'' depict $\gamma_D(x)$ and $\gamma_S(x)$, respectively.  In the lower figure, the dashed line in purple shows the approximated solution (\ref{eq:asym_sol_YB}) due to the decay of the hot relic, $Y_B / \epsilon_1 \sim Y_{\rm hot}=\frac{45}{2\pi}\cdot\frac{g_{X_1}}{h_{\rm eff}}$.}
 \label{fig:net_number}
\end{figure}

To see the concrete evolution dynamics, we consider the following interaction
\begin{eqnarray}
 \mathcal{L}_{\rm int}
  &=& -\sum_{a,i}y_{ai}\phi X_a\psi_i +({\rm h.c.})
\end{eqnarray}
with the Yukawa coupling $y_{ai}$ and the two-component spinors $X_a$ and $\psi_i$.  This interaction leads the concrete representation of the decay width and the scattering rate as
\begin{eqnarray}
 \Gamma_{X_1}
  &=& \tilde{\alpha} M_{X_1}, \\
 \Gamma_S
  &=& T\cdot \frac{8\tilde{\alpha}^2}{\pi g_{\psi}}\tilde{\gamma}_S(x)
\end{eqnarray}
where we denoted
\begin{eqnarray}
 \tilde{\alpha}
  &=& \sum_ig_{\psi_i}g_\phi\frac{|y_{1i}|^2}{32\pi}, \\
 \tilde{\gamma}_S
  &=& \frac{1}{8}\int_0^\infty dz \:  K_1(z)\left[2\left(\frac{z^4}{z^2+x^2} +\frac{x^2z^2}{z^2+2x^2} \ln\left(1+\frac{z^2}{x^2}\right)\right) \right. \nonumber \\
  & & \qquad \qquad +\frac{x^2z^4}{(z^2-x^2)^2+\tilde{\alpha}^2x^4}+2\left(z^2-x^2\ln\left(1+\frac{z^2}{x^2}\right)\right) \: \nonumber \\
  & & \qquad \qquad \left. + \frac{4x^2(z^2-x^2)}{(z^2-x^2)^2+\tilde{\alpha}^2x^4}\left(z^2-\left(z^2+x^2\right)\ln\left(1+\frac{z^2}{x^2}\right)\right)\right] \label{eq:gamma_S_tilde} \\
  &\sim & \left\{ \begin{array}{cl} 1 & (x\ll 1) \\ 8/x^2 & (x \gg 1) \end{array} \right.,
\end{eqnarray}
and $g_\psi\equiv \sum_ig_{\psi_i}=N_\psi g_{\psi_i}$.  Here $g_{\psi_i}$ and $g_\phi$ are the degrees of freedom of the chiral fermion $\psi_i$ and the scalar $\phi$, not including their anti-particle state.  The asymptotic behaviors for each reaction rate are governed by 
\begin{equation}
 \gamma_D(x) \:\: \propto \:\: \left\{ \begin{array}{cc} x^2 & (x\ll 1) \\ x^1 & (x\gg 1) \end{array} \right. , \qquad \gamma_S(x) \:\: \propto \:\: \left\{ \begin{array}{cc} ({\rm constant}) & (x\ll 1) \\ x^{-2} & (x\gg 1) \end{array}. \right. \label{eq:asymptotic_behavior}
\end{equation}
The actual behavior of $\gamma_D$ and $\gamma_S$ with concrete parameters are shown in the upper side of Figure \ref{fig:net_number}.  
The asymptotic behaviors at $x\ll 1$ and $x\gg 1$ are consistent with (\ref{eq:asymptotic_behavior}). 
The enhancement structures for each $\gamma_S$ seen around $x\sim \mathcal{O}(1)$ are induced by the resonant process through the on-shell $s$-channel shown in (\ref{eq:gamma_S_tilde}).

The lower side in Figure \ref{fig:net_number} shows the evolution of $Y_B/\epsilon_1$, which is the numerical result from the coupled equations (\ref{eq:eom_X_Y}) and (\ref{eq:eom_B_Y}).  The result shows that the heavier mass of X can generate more the net baryon number because the reaction rates are reduced for the heavier case, and hence the generated baryons can avoid the wash-out process.  Especially, the plot for $M_{X_1}=10^{16}$ GeV leads the close result to the hot relic approximation (\ref{eq:YB_inf_approx1}), whereas the plot for $M_{X_1}=10^{13}$ GeV shows the dumping by the wash-out effect at the late stage.  The milder decrease at the middle stage is caused by the decay of $X$ particles that supplies the net baryons to compensate for the wash-out effect.

Finally, the obtained yield of the net baryon number $Y_B$ should be compared with the current bound (\ref{eq:baryon_bound}), $Y_{B,{\rm now}}\sim 10^{-10}$.  Since the mean net baryon number can roughly be estimated by $\epsilon_1\sim \tilde{\alpha}^2\sin^2 \theta_{\rm CP}$ where $\theta_{\rm CP}$ is a CP phase in the considered model, one can obtain the constraint from the current observation as $Y_{B,{\rm now}}/\epsilon_1\gtrsim 10^{-10}/\tilde{\alpha}^2\sim 10^{-6}$, where we used $\tilde{\alpha}=0.01$.  Therefore, one can find that $M_{X_1}\gtrsim 10^{14}$ GeV is allowed by compared with the lower plot in Figure \ref{fig:net_number}.

\section{Summary}
In this paper we have demonstrated the derivation of the Boltzmann equation from the microscopic point of view with the quantum field theory, in which the transition probability has been constructed with the statistically averaged quantum states.  Although both results of the full and the integrated Boltzmann equation (\ref{eq:full_Boltzmann}) and (\ref{eq:integrated_Boltzmann}) are consistent with the well-known results, our derivation ensures that especially the full Boltzmann equation is widely applicable even in the non-equilibrium state since the derivation does not assume any distribution type nor the temperature of the system.  Especially the integrated Boltzmann equation (\ref{eq:integrated_Boltzmann}) is quite convenient and applicable for wide situations.  In the particular case that the kinetic equilibrium cannot be ensured, the coupled equations with the temperature parameter (\ref{eq:integrated_Boltzmann_neq1}) and (\ref{eq:integrated_Boltzmann_neq2}) are better for following the dynamics.

As the application examples of the (integrated) Boltzmann equation in cosmology, we have reviewed two cases, the relic abundance of the DM and the baryogenesis scenario.  For the former case, we have shown the Boltzmann equation and its analysis.  The analytic results (\ref{eq:xf}) and (\ref{eq:Y_inf}) are quite helpful for estimating the final relic abundance of the DM and its freeze-out epoch. For the latter case, we have derived the Boltzmann equation with a specific model and show the numerical analysis.  The final net baryon number can be estimated by the analytic result (\ref{eq:YB_inf_approx1}) in the case of the weakly interacting system, whereas that is strongly suppressed by the wash-out effect in the case of the strongly interacting system.


The Boltzmann equation is a powerful tool for following the evolution of the particle number or other thermal quantities, and thus it will be applied for many more situations in future and will open a new frontier of the current physics.  We hope this paper helps you to use the Boltzmann equation and its techniques thoughtfully.

\section*{Acknowledgment}
We thank Chengfeng Cai, Yi-Lei Tang, and Masato Yamanaka for useful discussions and comments. This work is supported in part by the National Natural Science Foundation of China under Grant No. 12275367, and the Sun Yat-Sen University Science Foundation.

\appendix

\section{Validity of the Maxwell-Boltzmann similarity approximation} \label{sec:derivation_MB_approx}
Although the approximation of the distribution function by the Maxwell Boltzmann {\it similarity} distribution is used well in many situations, such approximation is not always valid.  In this appendix, we show that the approximation is valid if the focusing species is in the kinetic equilibrium through interacting with the thermal bath.

Let us consider the situation of the particle number conserving process $a(k_1)+b(k_2)\leftrightarrow a(k_3)+b(k_4)$, where $a$ and $b$ denote the particle species.  If this process happens fast enough and the species $b$ maintains the thermal distribution, the condition of the detailed balance leads
\begin{eqnarray}
 0 &=& f_a(t,E_1)f_b^{\rm MB}(t,E_2)-f_a(t,E_3)f_b^{\rm MB}(t,E_4) \\
  &=& \left(\frac{f_a(t,E_1)}{f_a^{\rm MB}(t,E_1)}-\frac{f_a(t,E_3)}{f_a^{\rm MB}(t,E_3)}\right)f_a^{\rm MB}(t,E_1)f_b^{\rm MB}(t,E_2),
\end{eqnarray}
where we assumed the common temperature to the thermal bath and the energy conservation law: $f_a^{\rm MB}(t,E_1)f_b^{\rm MB}(t,E_2)=f_a^{\rm MB}(t,E_3)f_b^{\rm MB}(t,E_4)$.  Because the above relation must be satisfied by arbitrary energy, one can obtain
\begin{equation}
 \frac{f_a(t,E)}{f_a^{\rm MB}(t,E)} \:\: = \:\: C(t)
\end{equation}
where $C(t)$ is a function which is dependent on time but independent of the energy.  The function $C(t)$ can be determined by integrating over the momentum of $f_a(t,E)= C(t)f_a^{\rm MB}(t,E_1)$, i.e., $n_a(t)=C(t)n_a^{\rm MB}(t)$.  Finally, one can obtain the desired form of the distribution:
\begin{equation}
 f_a(t,E) \:\:=\:\: C(t)f_a^{\rm MB}(t,E) \:\:=\:\: \frac{n_a(t)}{n_a^{\rm MB}(t)}f_a^{\rm MB}(t,E).
\end{equation}

\section{Formulae for thermal average by Boltzmann-Maxwell distribution} \label{sec:formulae_TA}
In this section, we summarize the convenient formulae used in the various thermally averaged quantities  by the Maxwell-Boltzmann distribution, especially for number density, decay rate, and cross section.

\subsection{Number density and modified Bessel function}
The number density with the Maxwell-Boltzmann distribution is given by
\begin{eqnarray}
 n^{\rm MB}
  &=& g\int\frac{d^3k}{(2\pi)^3} f^{\rm MB} \\
  &=& \frac{g}{2\pi^2} m^2T K_2(m/T)e^{\mu/T} \label{eq:number_density_bessel}\\
  &=& g\times \left\{ \begin{array}{cc} \displaystyle \frac{1}{\pi^2}T^3e^{\mu/T}+\cdots & (T \gg m) \\ \displaystyle \left(\frac{mT}{2\pi}\right)^{3/2}e^{-(m-\mu)/T}\left(1+\frac{15T}{8m}+\cdots\right) & (T\ll m) \end{array} \right.,
\end{eqnarray}
where $K_n$ is the $n$-th order of the modified Bessel function given by
\begin{eqnarray}
 K_n(x)
  &=& \int_0^\infty d\theta \: e^{-x\cosh \theta}\cosh n\theta \\
  &=& \left\{ \begin{array}{cc} \displaystyle \frac{\Gamma(n)}{2}\left(\frac{2}{x}\right)^n+\cdots & (0<x \ll \sqrt{1+n}) \\ \displaystyle \sqrt{\frac{\pi}{2x}}\: e^{-x}\left(1+\frac{4n^2-1}{8x}+\cdots\right) & (x\gg 1 ) \end{array} \right. \label{eq:BesselK_asymptotic_exp}
\end{eqnarray}
Especially, the following relations are helpful in analysis:
\begin{eqnarray}
 K_n(x) &=& \frac{x}{2n}\left(K_{n+1}(x)-K_{n-1}(x)\right), \\
 \frac{d}{dx}\left(x^nK_n(x)\right)
  &=& -x^nK_{n-1}(x).
\end{eqnarray}

\subsection{Thermally averaged decay rate}
The rate defined in (\ref{eq:def_rate}) for the single initial species relates to the decay rate, $\mathcal{R}(A\rightarrow X,Y,\cdots)$ $=\frac{m_A}{2E_A}\Gamma_{A\rightarrow X,Y,\cdots}$, where
\begin{eqnarray}
 \Gamma_{A\rightarrow X,Y,\cdots}
  &=& \frac{1}{2m_A}\int\frac{d^3k_X}{(2\pi)^3}\frac{d^3k_Y}{(2\pi)^3}\cdots\frac{1}{2E_X2E_Y}\cdots   \nonumber \\
  & & \qquad \qquad \times (2\pi)^4\delta^4(k_A-k_X-k_Y-\cdots)\nonumber \\
  & & \qquad \qquad \times \frac{1}{g_A}\sum_{g_A,g_X,g_Y,\cdots}|\mathcal{M}(A\rightarrow X,Y,\cdots)|^2
\end{eqnarray}
is the partial width for the process $A\rightarrow X,Y,\cdots$.  The factor $m_A/E_A$ in $\mathcal{R}$ corresponds to the inverse Lorentz gamma factor describing the life-time dilation.  The thermal average of the rate is given by
\begin{eqnarray}
 \langle \mathcal{R}(A\rightarrow X,Y,\cdots)\rangle
  &=& \frac{1}{2}\Gamma_{A\rightarrow X,Y,\cdots} \left<\frac{m_A}{E_A}\right>, \\
 \left<\frac{m_A}{E_A}\right>
  &=& \frac{g_A}{n_A^{\rm MB}}\int\frac{d^3k_A}{(2\pi)^3} \frac{m_A}{E_A}f_A^{\rm MB} \\
  &=& \frac{K_1(m_A/T)}{K_2(m_A/T)} \\
  &=& \left\{ \begin{array}{cc} \displaystyle \frac{m_A}{2T}+\cdots & (T \gg m_A) \\ \displaystyle 1-\frac{3T}{2m_A}+\cdots & (T\ll m_A) \end{array} \right..
\end{eqnarray}

\subsection{Thermal averaged cross section}
The rate averaged by the initial 2-species relates to the scattering rate,
\begin{eqnarray}
 \mathcal{R}(A,B\rightarrow X,Y,\cdots)
  &=& \sigma v \\
  &=& \frac{1}{2E_A2E_B}\int\frac{d^3k_X}{(2\pi)^3}\frac{d^3k_Y}{(2\pi)^3}\cdots\frac{1}{2E_X2E_Y}\cdots   \nonumber \\
  & & \quad \times (2\pi)^4\delta^4(k_A+k_B-k_X-k_Y-\cdots)\nonumber \\
  & & \quad \times \frac{1}{g_Ag_B}\sum_{g_A,g_B,g_X,g_Y,\cdots}|\mathcal{M}(A\rightarrow X,Y,\cdots)|^2,
\end{eqnarray}
where $\sigma=\sigma(s)$ is the cross section for the process $A,B\rightarrow X,Y,\cdots$ dependent on the Mandelstam variable $s=(k_A+k_B)^2$, and $v$ is the M\o ller velocity
\begin{equation}
 v = \frac{\sqrt{(k_A\cdot k_B)^2-m_A^2m_B^2}}{E_AE_B}=\frac{\sqrt{(s-(m_A+m_B)^2)(s-(m_A-m_B)^2)}}{2E_AE_B}.
\end{equation}
The thermal average of the rate can be obtained by
\begin{eqnarray}
 \langle \mathcal{R}(A,B\rightarrow X,Y,\cdots)\rangle
  &=& \langle\sigma v\rangle \\
  &=& \frac{g_Ag_B}{n_A^{\rm MB}n_B^{\rm MB}}\int\frac{d^3k_A}{(2\pi)^3}\frac{d^3k_B}{(2\pi)^3} \: \sigma v\cdot f_A^{\rm MB} f_B^{\rm MB} \\
  &=&  \frac{g_Ag_B}{n_A^{\rm MB}n_B^{\rm MB}}\int_0^\infty d|\vec{k}_A| \: d|\vec{k}_B|\int_0^\pi d\theta \cdot \frac{1}{4\pi^2}\frac{|\vec{k}_A|^2|\vec{k}_B|^2\sin\theta}{E_AE_B} \nonumber \\
  & & \qquad \qquad \times \sigma(s) \cdot \sqrt{(s-(m_A+m_B)^2)(s-(m_A-m_B)^2)} \nonumber \\
  & & \qquad \qquad \times \exp\left[-\frac{E_A+E_B}{T}+\frac{\mu_A+\mu_B}{T}\right], \label{eq:sigma_v_integ1}
\end{eqnarray}
where the integral variable $\theta$ denotes the angle between $\vec{k}_A$ and $\vec{k}_B$, i.e., $\vec{k}_A\cdot\vec{k}_B=|\vec{k}_A||\vec{k}_B|\cos\theta$.

In order to perform the integral in (\ref{eq:sigma_v_integ1}), it is convenient to change the integral variables $(|\vec{k}_A|, |\vec{k}_B|, \theta)$ to $(E_+, E_-, s)$, where $E_\pm\equiv E_A\pm E_B$ \cite{Gondolo:1990dk}.  The Jacobian is given by
\begin{eqnarray}
 \left|\frac{\partial(|\vec{k}_A|,|\vec{k}_B|,\theta)}{\partial(E_+,E_-,s)}\right|
  &=& \frac{E_AE_B}{4|\vec{k}_A|^2|\vec{k}_B|^2\sin\theta}.
\end{eqnarray}
The integral region can be obtained from the expression of the Mandelstam variable,
\begin{equation}
 s = m_A^2+m_B^2+2\left(E_AE_B+|\vec{k}_A||\vec{k}_B|\cos\theta\right),
\end{equation}
which leads
\begin{equation}
 (s-m_A^2-m_B^2-2E_AE_B)^2 \leq 4|\vec{k}_A|^2|\vec{k}_B|^2=4(E_A^2-m_A^2)(E_B^2-m_B^2).
\end{equation}
The above inequality is equivalent to
\begin{equation}
 \left(E_--\frac{m_A^2-m_B^2}{s}E_+\right)^2\leq (E_+^2-s)\left(1-\frac{(m_A+m_B)^2}{s}\right)\left(1-\frac{(m_A-m_B)^2}{s}\right)
\end{equation}
Therefore, the integral region can be obtained as
\begin{equation}
 e_-\leq E_-\leq e_+, 
\end{equation}
\begin{equation}
 E_+ \geq \sqrt{s}, 
\end{equation}
\begin{equation}
 s \geq (m_A+m_B)^2,
\end{equation}
where
\begin{equation}
 e_\pm\equiv \frac{m_A^2-m_B^2}{s}E_+\pm\sqrt{(E_+^2-s)\left(1-\frac{(m_A+m_B)^2}{s}\right)\left(1-\frac{(m_A-m_B)^2}{s}\right)}.
\end{equation}
Using the above results, the integral (\ref{eq:sigma_v_integ1}) can be performed as
\begin{eqnarray}
 \langle\sigma v\rangle
  &=& \frac{g_Ag_B}{n_A^{\rm MB}n_B^{\rm MB}}\frac{1}{2(2\pi)^4}e^{(\mu_A+\mu_B)/T} \nonumber \\
  & & \times \int_{(m_A+m_B)^2}^\infty ds \cdot \sigma(s) \cdot (s-(m_A+m_B)^2)(s-(m_A-m_B)^2)\nonumber \\
 & & \qquad \qquad \times \frac{T}{\sqrt{s}}K_1(\sqrt{s}/T) \\
 &=& \frac{1}{4m_A^2m_B^2T} \int_{m_A+m_B}^\infty d\sqrt{s} \cdot \sigma(s) \cdot (s-(m_A+m_B)^2)\nonumber \\
 & & \qquad \qquad \qquad \times(s-(m_A-m_B)^2) \cdot \frac{K_1(\sqrt{s}/T)}{K_2(m_A/T)K_2(m_B/T)}, \label{eq:thermal_averaged_cs1}
\end{eqnarray}
where we used the representation of the number density (\ref{eq:number_density_bessel}).

Especially in the case of the non-relativistic limit, $m_A, m_B \gg T$, it is convenient to use the representation\footnote{The Lorentz-invariant ``velocity'' $v_{\rm NR}$ behaves as $v_{\rm NR}\sim\left|\frac{\vec{k}_A}{m_A}-\frac{\vec{k}_B}{m_B}\right|$ at the non-relativistic limit.  Note that $\displaystyle \lim_{v_{\rm NR}\rightarrow 0}\widetilde{\sigma v}$ remains non-zero ($s$-wave contribution) in general.}
\begin{eqnarray}
 \widetilde{\sigma v}(s)
  &\equiv& \sigma(s)\cdot v_{\rm NR}(s), \qquad v_{\rm NR}(s) \:\: \equiv \:\: \sqrt{\frac{s-(m_A+m_B)^2}{m_Am_B}}
\end{eqnarray}
and the replacement of the integral variable $s$ to $y$ defined by
\begin{eqnarray}
 \sqrt{s} &=& m_A+m_B+Ty.
\end{eqnarray}
Since the integral parameter $y$ corresponds to $\vec{k}^2/mT$ naively, we can expect that the significant integral interval is on $y\lesssim \mathcal{O}(1)$.
Then (\ref{eq:thermal_averaged_cs1}) can be approximated as
\begin{eqnarray}
 \langle\sigma v\rangle
  &\sim& \frac{2}{\sqrt{\pi}}\left(1-\frac{15T}{8m_A}-\frac{15T}{8m_B}+\frac{3T}{8(m_A+m_B)}+\cdots\right) \nonumber \\
  & & \times \int_0^\infty dy \cdot \left((\widetilde{\sigma v})_0+Ty\cdot (\widetilde{\sigma v})'_0+ \cdots\right)\nonumber \\
  & & \qquad \qquad \times e^{-y}\cdot \sqrt{y}\left(1+\frac{Ty}{2m_A}+\frac{Ty}{2m_B}-\frac{Ty}{4(m_A+m_B)}+\cdots\right) \\
  &=& (\widetilde{\sigma v})_0 +\frac{3}{2}T\left[-\frac{3}{4}\left(\frac{1}{m_A}+\frac{1}{m_B}\right)(\widetilde{\sigma v})_0+ (\widetilde{\sigma v})'_0\right]+ \mathcal{O}(T^2), \label{eq:thermal_averaged_cs2}
\end{eqnarray}
where we used the asymptotic expansion (\ref{eq:BesselK_asymptotic_exp}) and the Taylor series around $\sqrt{s}=m_A+m_B$,
\begin{equation}
 \widetilde{\sigma v} \:\: = \:\: (\widetilde{\sigma v})_0+(\sqrt{s}-m_A-m_B)\cdot (\widetilde{\sigma v})'_0+\cdots
\end{equation}
\begin{eqnarray}
 (\widetilde{\sigma v})_0 \:\: \equiv \:\: \widetilde{\sigma v}(\sqrt{s}=m_A+m_B), \qquad (\widetilde{\sigma v})'_0 \:\: \equiv \:\: \left.\frac{d\:\widetilde{\sigma v}}{d\sqrt{s}}\right|_{\sqrt{s}=m_A+m_B}.
\end{eqnarray}

\section{Derivation of the Boltzmann equations in baryogenesis scenario} \label{sec:derivation_of_boltzmann_eq}
In this section, we demonstrate the derivation of the Boltzmann equation in the baryogenesis scenario with the processes listed in Table \ref{tab:matter_contents}.  
Indeed, the straightforward derivation of the Boltzmann equations leads to the over-counting problem in the amplitudes.  For example, once a contribution of the decay/inverse-decay process $X\leftrightarrow \psi,\phi$ is included in the Boltzmann equation, the straightforward contribution from the scattering process $\bar{\psi},\bar{\phi}\leftrightarrow\psi,\phi$ is over-counted because such process can be divided into $\bar{\psi},\bar{\phi}\leftrightarrow X$ and $X\leftrightarrow\psi,\phi$ if the intermediate state $X$ is on-shell.  Therefore, in general, one must regard the straightforward contribution in the scattering processes as the subtracted state of the real intermediated state (RIS) from the full contribution \cite{Kolb:1990vq,Buchmuller:2004nz}:
\begin{eqnarray*}
 |\mathcal{M}|_{\rm Boltzmann \ eq.}^2
  &=& |\mathcal{M}|_{\rm subtracted}^2 \:\: \equiv \:\: |\mathcal{M}|_{\rm full}^2-|\mathcal{M}|_{\rm RIS}^2.
\end{eqnarray*}
In a case of the scattering process $\bar{\psi},\bar{\phi}\rightarrow\psi,\phi$, the full amplitude part can be represented as
\begin{eqnarray}
 i\mathcal{M}(\bar{\psi},\bar{\phi}\rightarrow\psi,\phi)_{\rm full}
  &\sim& i\mathcal{M}(X\rightarrow\psi,\phi)\cdot \frac{i}{s-M_X^2+iM_X\Gamma_X}\cdot i\mathcal{M}(\bar{\psi},\bar{\phi}\rightarrow X) \label{eq:full_amplitude}
\end{eqnarray}
where $s$ is the Mandelstam variable, $M_X$ and $\Gamma_X$ are $X$'s mass and total decay width, respectively.  On the other hand, the RIS part can be evaluated as the limit of the narrow width by
\begin{eqnarray}
 \left|\mathcal{M}(\bar{\psi},\bar{\phi}\rightarrow\psi,\phi)\right|_{\rm RIS}^2
  &=& \lim_{\Gamma_X\rightarrow 0}\left|\mathcal{M}(\bar{\psi},\bar{\phi}\rightarrow\psi,\phi)\right|_{\rm full}^2 \\
  &=& \lim_{\Gamma_X\rightarrow 0}|\mathcal{M}(X\rightarrow\psi,\phi)|^2\frac{1}{(s-M_X^2)^2+(M_X\Gamma_X)^2}|\mathcal{M}(\bar{\psi},\bar{\phi}\rightarrow X)|^2 \nonumber \\
  &\sim& |\mathcal{M}(X\rightarrow\psi,\phi)|^2\cdot \frac{\pi}{M_X\Gamma_X}\delta(s-M_X^2)\cdot |\mathcal{M}(\bar{\psi},\bar{\phi}\rightarrow X)|^2. \label{eq:RIS_appro}
\end{eqnarray}
In the last line, the narrow width approximation is applied.  Since the contribution of both amplitudes in (\ref{eq:RIS_appro}) is the order of $\Gamma_X$, the RIS contribution is also the order of $\Gamma_X$ in total. Therefore, RIS part in the scattering process contributes to the decay/inverse-decay process.
 
Taking into account the above notice, we derive the Boltzmann equation.  For simplicity, we suppose that only a single flavour $X_1$ affects to the evolution of the net baryon number.  Because of no scattering processes associated with $X_a$,  the equation governing $n_{X_a}$ is simply written as
\begin{eqnarray}
 \dot{n}_{X_1}+3Hn_{X_1}
  &=& \int\frac{d^3k_{X_1}}{(2\pi)^3}\frac{d^3k_{\psi_i}}{(2\pi)^3}\frac{d^3k_\phi}{(2\pi)^3}\frac{1}{2E{X_1}2E_{\psi_i}2E_\phi}(2\pi)^4\delta^4(k_{X_1}-k_{\psi_i}-k_\phi) \nonumber \\
  & & \times \frac{1}{g_{X_1}}\sum_{g_{X_1},g_{\psi_i},g_\phi}\left[ -f_{X_1}|\mathcal{M}(X_1\rightarrow\psi_i\phi)|^2+f_{\psi_i}f_\phi|\mathcal{M}(\psi_i\phi\rightarrow X_1)|^2 \right.\nonumber \\
  & & \qquad \qquad \qquad \left.-f_{X_1}|\mathcal{M}(X_1\rightarrow\bar{\psi}_i\bar{\phi})|^2+f_{\bar{\psi}_i}f_{\bar{\phi}}|\mathcal{M}(\bar{\psi}_i\bar{\phi}\rightarrow X_1)|^2 \right] \nonumber \\
  & & +\cdots \\
  &=& -\sum_i\langle \Gamma_{X_1}\rangle(n_{X_1}-n_{X_1}^{\rm MB})+\cdots \label{eq:Boltzmann_eq_X}
\end{eqnarray}
where
\begin{eqnarray}
 \Gamma_{X_1}
  &=& \Gamma_{X_1\rightarrow \psi_i\phi}+\Gamma_{X_1\rightarrow \bar{\psi}_i\bar{\phi}}+\cdots
\end{eqnarray}
is the total decay width of $X_1$ and
\begin{eqnarray}
 \langle\Gamma_{X_1}\rangle
  &\equiv& \frac{1}{n_{X_1}^{\rm MB}}\sum_{g_{X_1}}\int\frac{d^3k_{X_1}}{(2\pi)^3}\frac{M_{X_1}}{E_{X_1}}\Gamma_{X_1}f_{X_1}^{\rm MB} \\
  &=& \Gamma_{X_1}\cdot\frac{K_1(M_{X_1}/T)}{K_2(M_{X_1}/T)} \:\:\sim\:\: \Gamma_{X_1}\times \left\{ \begin{array}{cc} M_{X_1}/2T & (M_{X_1}\ll T) \\ 1 & (M_{X_1}\gg T) \end{array} \right.
\end{eqnarray}
is the thermally averaged width, $K_n(x)$ is the modified Bessel function.  To derive (\ref{eq:Boltzmann_eq_X}), we assumed the universal distributions for $\psi_i$ ($f_{\psi_i}\sim \frac{1}{N_\psi}f_\psi, f_\psi\equiv\sum_if_{\psi_i}$) and ignored the chemical potentials in the thermal distributions ($f_\psi^{\rm MB}=f_{\bar{\psi}}^{\rm MB}, \ f_\phi^{\rm MB}=f_{\bar{\phi}}^{\rm MB}$).  Besides, we assumed $\phi$ is always in the thermal equilibrium ($f_\phi=f_\phi^{\rm MB}$). On the other hand, $\psi_i$'s equation should be derived with the consideration of the subtracted state in some scattering processes to avoid the over-counting of the decay/inverse-decay processes:
\begin{eqnarray}
 \dot{n}_\psi+3Hn_\psi
  &=& \sum_i\int\frac{d^3k_{X_1}}{(2\pi)^3}\frac{d^3k_{\psi_i}}{(2\pi)^3}\frac{d^3k_\phi}{(2\pi)^3}\frac{1}{2E{X_1}2E_{\psi_i}2E_\phi}(2\pi)^4\delta^4(k_{X_1}-k_{\psi_i}-k_\phi) \nonumber \\
  & & \quad \times \sum_{g_{X_1},g_{\psi_i},g_\phi}\left[ f_{X_1}|\mathcal{M}(X_1\rightarrow\psi_i\phi)|^2-f_{\psi_i}f_\phi |\mathcal{M}(\psi_i\phi\rightarrow X_1)|^2 \right] \nonumber \\
  & & +\sum_{i,j}\int\frac{d^3k_{\psi_i}}{(2\pi)^3}\frac{d^3k_{\psi_j}}{(2\pi)^3}\frac{d^3k_{\phi_1}}{(2\pi)^3}\frac{d^3k_{\phi_2}}{(2\pi)^3}\frac{1}{2E_{\psi_i}2E_{\psi_j}2E_{\phi_1}2E_{\phi_2}}\sum_{g_{\psi_i},g_{\psi_j},g_{\phi_1},g_{\phi_2}} \nonumber \\
  & & \quad \times \left[ (2\pi)^4\delta^4(k_{\psi_i}+k_{\psi_j}-k_{\phi_1}-k_{\phi_2}) \right. \nonumber \\
  & & \quad \qquad \times \left( -f_{\psi_i}f_{\psi_j}|\mathcal{M}(\psi_i\psi_j\rightarrow\bar{\phi}_1\bar{\phi}_2)|^2+f_{\bar{\phi}_1}f_{\bar{\phi}_2}|\mathcal{M}(\bar{\phi}_1\bar{\phi}_2\rightarrow\psi_i\psi_j)|^2 \right) \nonumber \\
  & & \qquad +(2\pi)^4\delta^4(k_{\psi_i}+k_{\phi_1}-k_{\psi_j}-k_{\phi_2}) \nonumber \\
  & & \quad \qquad \times \left. \left( -f_{\psi_i}f_{\phi_1}|\mathcal{M}(\psi_i\phi_1\rightarrow\bar{\psi}_j\bar{\phi}_2)|_{\rm sub}^2+f_{\bar{\psi}_j}f_{\bar{\phi}_2}|\mathcal{M}(\bar{\psi}_j\bar{\phi}_2\rightarrow\psi_i\phi_1)|_{\rm sub}^2 \right)\right] \nonumber \\
  & & +\cdots \\
  &=& \left( n_{X_1}\langle\Gamma_{X_1\rightarrow\psi\phi}\rangle-n_{X_1}^{\rm MB}\frac{n_{\psi}}{n_{\psi}^{\rm MB}} \langle\Gamma_{X_1\rightarrow \bar{\psi}\bar{\phi}}\rangle \right) \nonumber \\
  & & -(n_{\psi})^2\langle\sigma_{\psi\psi\rightarrow\bar{\phi}\bar{\phi}}v\rangle+(n_\psi^{\rm MB})^2\langle\sigma_{\bar{\psi}\bar{\psi}\rightarrow\phi\phi}v\rangle \nonumber \\
  & & -n_\phi^{\rm MB}\left( n_{\psi}\langle\sigma_{\psi\phi\rightarrow\bar{\psi}\bar{\phi}}v\rangle-n_{\bar{\psi}}\langle\sigma_{\bar{\psi}\bar{\phi}\rightarrow\psi\phi}v\rangle \right) \nonumber \\
  & & +\left(\frac{n_{\psi}}{n_{\psi}^{\rm MB}}\frac{\langle(\Gamma_{X_1\rightarrow\bar{\psi}\bar{\phi}})^2\rangle}{\Gamma_{X_1}}- \frac{n_{\bar{\psi}}}{n_{\psi}^{\rm MB}}\frac{\langle(\Gamma_{X_1\rightarrow\psi\phi})^2\rangle}{\Gamma_{X_1}}\right)n_{X_1}^{\rm MB} \nonumber \\
  & & +\cdots \label{eq:boltzmann_eq_psi}
\end{eqnarray}
where we used the notations $n_\psi\equiv \sum_i n_{\psi_i}$, $\Gamma_{X_a\rightarrow\psi\phi}\equiv\sum_i\Gamma_{X_a\rightarrow\psi_i\phi}$, and
\begin{eqnarray}
 \langle\sigma_{\psi\phi\rightarrow\bar{\psi}\bar{\phi}}v\rangle
  &\equiv& \frac{1}{n_\psi^{\rm MB}n_\phi^{\rm MB}}\cdot g_{\psi}g_\phi\int\frac{d^3k_{\psi_i}}{(2\pi)^3}\frac{d^3k_\phi}{(2\pi)^3}\sigma_{\psi\phi\rightarrow\bar{\psi}\bar{\phi}}v\cdot f_{\psi_i}^{\rm MB}f_\phi^{\rm MB}\\
 \langle\sigma_{\psi\psi\rightarrow\bar{\phi}\bar{\phi}}v\rangle
  &\equiv& \frac{1}{(n_\psi^{\rm MB})^2}\cdot g_\psi^2 \int\frac{d^3k_{\psi_i}}{(2\pi)^3}\frac{d^3k_{\psi_j}}{(2\pi)^3}\sigma_{\psi\psi\rightarrow\bar{\phi}\bar{\phi}}v\cdot f_{\psi_i}^{\rm MB}f_{\psi_j}^{\rm MB}
\end{eqnarray}
with $g_\psi\equiv \sum_ig_{\psi_i}$ are the thermally averaged cross sections.  The fourth line in (\ref{eq:boltzmann_eq_psi}) corresponds to the RIS contribution that makes the thermal balance to the first line, while the processes in the third line includes the resonant structure as seen in (\ref{eq:full_amplitude}). With the expression of the net baryon number density $n_B=b(n_\psi-n_{\bar{\psi}})$, one can finally obtain the equation for the net baryons using (\ref{eq:boltzmann_eq_psi}) as
\begin{eqnarray}
 \dot{n}_B+3Hn_B
  &=& \epsilon_1\langle\Gamma_{X_1}\rangle\left(n_{X_1}- n_{X_1}^{\rm MB}\right) \nonumber \\
  & &  -n_\phi^{\rm MB} \left[2bn_\psi^{\rm MB}\left(\langle\sigma_{\psi\phi\rightarrow \bar{\psi}\bar{\phi}}v\rangle-\langle\sigma_{\bar{\psi}\bar{\phi}\rightarrow \psi\phi}v\rangle\right) \right. \nonumber \\
  & & \qquad \qquad \left. + n_B\left(\langle\sigma_{\psi_i\phi\rightarrow \bar{\psi}\bar{\phi}}v\rangle+\langle\sigma_{\bar{\psi}_i\bar{\phi}\rightarrow \psi\phi}v\rangle\right)\right] \nonumber \\
  & &  -n_\psi^{\rm MB}\left[2bn_\psi^{\rm MB}\left(\langle\sigma_{\psi\psi\rightarrow \bar{\phi}\bar{\phi}}v\rangle-\langle\sigma_{\bar{\psi}\bar{\psi}\rightarrow \phi\phi}v\rangle\right) \right. \nonumber \\
  & & \qquad \qquad \left. + n_B\left(\langle\sigma_{\psi\psi\rightarrow \bar{\phi}\bar{\phi}}v\rangle+\langle\sigma_{\bar{\psi}\bar{\psi}\rightarrow \phi\phi}v\rangle\right)\right] \nonumber \\
  & &  +\cdots. \label{eq:EOM_net_baryon}
\end{eqnarray}
where $\epsilon_1$ is the mean net number defined in (\ref{eq:mean_net_baryon2}), and we used the approximation
\begin{equation}
 n_\psi+n_{\bar{\psi}} \ \sim \ 2n_\psi^{\rm MB} \ \gg \ |n_B| \ = \ b|n_\psi-n_{\bar{\psi}}|.
\end{equation}
Note that the tree level contribution of cross sections and their anti-state are same in general. Therefore, the terms in second and the fourth lines in (\ref{eq:EOM_net_baryon}) are cancelled in the leading order, respectively.

\end{document}